\documentclass[aps,prd,twocolumn,showpacs,10pt,superscriptaddress,preprintnumbers,nofootinbib,floatfix]{revtex4-1}
\usepackage[utf8]{inputenc}
\usepackage{amsmath,amssymb,bm,slashed,braket}
\usepackage{graphicx}
\usepackage{epstopdf}
\usepackage[dvipsnames,table]{xcolor}
\usepackage[normalem]{ulem}
\usepackage{colortbl}
\usepackage{multirow, booktabs}
\usepackage{makecell}
\usepackage{hhline}
\usepackage{dsfont}
\usepackage{verbatim}
\usepackage[colorlinks=true,
  linkcolor=blue,
  urlcolor=blue,
  citecolor=purple,
  bookmarks=true,
  bookmarksnumbered=true,
  breaklinks=true,
  pdfpagemode=Fullscreen,
pdfstartview=FitBH]{hyperref}

\allowdisplaybreaks[4]
\usepackage[capitalise]{cleveref}
\usepackage{stackengine}
\usepackage{orcidlink}
\usepackage{ctable}
\usepackage{pifont}
\usepackage{cancel}
\newcommand{\muep}{\mu^-\!\to e^+}
\newcommand{\cL}{\mathcal L}
\newcommand{\cO}{\mathcal O}
\newcommand{\cM}{\mathcal M}
\newcommand{\dd}{\mathrm d}
\newcommand{\eeff}{E_{\mathrm{eff}}}
\newcommand{\peff}{p_{\mathrm{eff}}}
\newcommand{\bpeff}{\mathbf p_{\mathrm{eff}}}
\newcommand{\bq}{\mathbf q}
\newcommand{\bp}{\mathbf p}
\newcommand{\bK}{\mathbf K}
\newcommand{\bB}{\mathbf B}
\newcommand{\bW}{\mathbf W}
\newcommand{\bL}{\mathbf L}
\newcommand{\br}{\mathbf r}
\newcommand{\bR}{\mathbf R}
\newcommand{\bx}{\mathbf x}
\newcommand{\bsigma}{\boldsymbol{\sigma}}

\begin{document}
\raggedbottom

\title{Muon-to-positron conversion from
local dimension-nine operators: pion-pole-improved nuclear multipoles}

\author{Yi Liao\,\orcidlink{0000-0002-1009-5483}}
\email{liaoy@m.scnu.edu.cn}
\author{Hao-Lin Wang\,\orcidlink{0000-0002-2803-5657}}
\email{whaolin@m.scnu.edu.cn}
\affiliation{State Key Laboratory of Nuclear Physics and Technology,
  Institute of Quantum Matter, South China Normal University,
Guangzhou 510006, China}
\affiliation{Guangdong Basic Research Center of Excellence for Structure and
  Fundamental Interactions of Matter, Guangdong Provincial Key Laboratory of
Nuclear Science, Guangzhou 510006, China}

\begin{abstract}
We begin a systematic study of neutrinoless muon-to-positron conversion, $\mu^-+(A,Z)\to e^++(A,Z-2)$, which violates lepton number by two units and changes charged-lepton flavor. This first paper in a series constructs nuclear operators for local dimension-nine interactions containing a scalar or pseudoscalar lepton bilinear and two contracted vector or axial quark currents. The covariant nucleon currents are built with all six vector and axial form factors. For the working nuclear operator we neglect second-class currents and adopt a pion-pole-improved reduction: non-pole recoil terms generated by the displayed one-body currents are retained through second order in inverse nucleon mass, whereas the pion-pole longitudinal axial current is kept intact before the two currents are contracted. We preserve the two independent momentum transfers at the nucleon vertices and the Coulomb-shifted spatial phase of the outgoing positron. Starting from the spacetime transition element, we show explicitly how the interaction-coordinate integral fixes the sum, but not the sharing, of the two current momenta. Unequal form-factor arguments then generate relative-coordinate multipoles absent in the usually used common-momentum approximation. We give an exact recoupling into two-body nuclear tensors and specialize it to the phenomenologically important $0^+\to0^+$ transition. In this case the external nuclear rank is zero and all magnetic sums collapse, although finite-momentum internal orbital and spin tensors remain.  The result is expressed in a form directly contractible with charge-changing two-body transition densities. General six-form-factor currents, arbitrary nuclear angular momenta, and power-counting qualifications are collected in appendices.
\end{abstract}

\maketitle

\section{Introduction}
\label{sec:introduction}

The origin of light neutrino mass and the possible violation of lepton number are central questions in particle and nuclear physics.  The Sakharov conditions frame the dynamical origin of the cosmic matter-antimatter asymmetry, and leptogenesis realizes them through lepton-number-violating (LNV) dynamics~\cite{Sakharov1967,FukugitaYanagida1986}. Majorana masses provide an economical explanation for the observed smallness of neutrino masses while offering an origin of LNV. In particular, the type-I seesaw mechanism~\cite{Minkowski1977,Yanagida1979,GellMannRamondSlansky1979,Glashow1980,MohapatraSenjanovic1980} provides a concrete ultraviolet realization of light Majorana neutrino mass. Neutrinoless double-beta decay ($0\nu\beta\beta$) is the most sensitive established probe of interactions that violate lepton number by two units ($\Delta L=2$)~\cite{SchechterValle1982,EngelMenendez2017,Dolinski2019}, but it probes primarily the electron-flavor sector.

Above the electroweak scale, LNV interactions can be organized in the standard model effective field theory (SMEFT) independent of their ultraviolate mechanism. The unique dimension-five Weinberg operator generates a Majorana neutrino mass after electroweak symmetry breaking~\cite{Weinberg1979}, while higher-dimensional operators encode additional long- and short-distance sources of LNV~\cite{BabuLeung2001,deGouveaJenkins2008}. Complete classifications of effective operators have been developed through dimension nine in SMEFT and in the low-energy effective field theory (LEFT)~\cite{Lehman2014,LiaoMa2016Dim7,LiaoMa2019Dim7,LiaoMa2020,Li:2020xlh,LiaoMaWang2020,LiEtAl2021}. Local
four-quark--two-lepton interactions with $\Delta L=2$ first occur at dimension nine in LEFT, with Wilson coefficients that inherit electroweak matching, QCD running, flavor, and operator-basis dependence~\cite{Graesser2017,LiaoEtAl2020}.

The short-range $0\nu\beta\beta$ provides a developed analogue to muon-to-positron conversion studied in this work.
Effective-field-theory (EFT) analyses can connect systematically dimension-nine (dim-9) to hadronic interactions~\cite{Prezeau2003,Cirigliano2018Master}. And detailed nucleon-current, nuclear-matrix-element, and phase-space reductions have also been derived for the short-range operator basis~\cite{Graf2018,DeppischGraf2020}. More recently, ab initio short-range matrix elements have become available for several principal candidate isotopes~\cite{ToddEtAl2026}. Representative modern many-body evaluations use beyond-mean-field covariant density functional theory, ab initio methods, projected shell-model methods, and deformed quasiparticle random-phase approximations~\cite{Yao2015,WirthYaoHergert2021,DingLiYao2024,WangZhaoMeng2021,FangFaesslerSimkovic2018}.
In a nonrelativistic chiral-EFT treatment of the light-neutrino mechanism, iteration of the long-range potential between nonperturbative two-nucleon states requires a local counterterm already at the leading order to absorb regulator dependence~\cite{Cirigliano2018Contact,Cirigliano2021Complete}. Yet a relativistic chiral-EFT analysis instead finds that no unknown contact interaction is required through next-to-leading order~\cite{YangZhao2025}. These differing conclusions concern the EFT organization of short-distance two-nucleon contributions. In this work we instead will adopt the impulse approximation while leaving a complete EFT analysis for the future work.

The process studied here is
\begin{equation}
  \mu^-+(A,Z)\longrightarrow e^++(A,Z-2),
  \label{eq:process}
\end{equation}
where $A,~Z$ are the mass and charge numbers of the parent nucleus. It violates lepton number by two units and changes charged-lepton flavor. This article is the first in our systematic program on $\muep$ conversion.
Here we will establish the spacetime, lepton-wave, single-nucleon-current, and nuclear-multipole framework for local dim-9 interactions; other mechanisms and quantitative many-body applications will be treated in subsequent work. So far, direct searches with titanium targets have constrained both conversion to the ground state and excited states of the daughter nucleus~\cite{Dohmen1993,Kaulard1998}. Future Mu2e and COMET experiments motivate renewed attention to the positron channel and to its radiative-muon-capture background~\cite{Mu2eTDR,COMETTDR,LeeMacKenzie2022}. Existing theory studies include light- and heavy-Majorana-neutrino exchange, effective-operator analyses, various
short-range phenomenology, and evaluation of nuclear matrix elements using shell-model and quasiparticle
random-phase-approximation~\cite{Simkovic2001,Divari2002,Domin2004,
GeibMerleZuber2017,Berryman2017,Geib:2016daa}.

The previous study most directly relevant to nuclear calculation is that of Domin \textit{et al.}, who evaluated light- and heavy-neutrino mechanisms for $^{48}$Ti in the renormalized proton--neutron quasiparticle random-phase approximation and used a nuclear average of the bound-muon wave function~\cite{Domin2004}. Geib and Merle later treated short-range operators, but adopted a common momentum in the two nucleon form factors~\cite{Geib:2016daa}. An overall Fermi-function correction cannot also retain the Coulomb-modified spatial phase of the emitted positron. These approximations are particularly consequential because the positron momentum and the inverse nuclear radius are comparable.

We will develop the nuclear reduction for a restricted but interesting set of local dim-9 operators: a scalar or pseudoscalar lepton bilinear multiplying two contracted vector or two contracted axial quark currents. The complete on-shell nucleon currents contain three form factors in each Lorentz channel. Carrying every one of them through the main analysis would obscure the physically relevant result.  We therefore will state the general result and then adopt two controlled specializations. First, the second-class induced-scalar and induced-tensor form factors are set to zero. Second, the remaining current is reduced with a pion-pole-improved (PI) prescription, retaining weak
magnetism and the complete spin-longitudinal pion-pole current. The general six-form-factor and arbitrary-$J_i\to J_f$ constructions are given in the appendices.

The Lorentz structures that we will focus on are among the simplest dim-9 operators and are physically well motivated by their close connection to the short-range operators studied in neutrinoless double-beta decay. The method developed below is not tied to this subset: after replacing the elementary lepton and nucleon-current tensors, our approach on the momentum routing, form-factor expansion, angular recoupling, and nuclear-density contraction generalizes straightforwardly to other local dim-9 operators.

Our main result is the exact PI two-body operator for a $0^+\to0^+$ nuclear transition. The scalar selection rule removes the external angular-momentum and magnetic sums, but does not set the internal orbital tensors to zero. We will also retain the two independent current momenta and use an effective positron plane wave that carries both the Coulomb normalization and the shifted spatial momentum. The final expression can be contracted directly with charge-changing two-body transition densities (TBTDs).

\section{From the local interaction to the spatial matrix element}
\label{sec:interaction-rate}

At a hadronic renormalization scale, we write the local short-range
interaction as
\begin{align}
  \cL_{\Delta L=2}^{(9)}
  &={}\frac{G_F^2}{2m_p}\sum_I C_I(\mu_{\rm had})\cO_I^{(9)}
  +\mathrm{H.c.},
  \nonumber\\[-2pt]
  \cO_I^{(9)}&=L_IJ_IJ'_I .
  \label{eq:general-lagrangian}
\end{align}
The coefficient $C_I$ is dimensionless in this convention. The label $I$
includes the Lorentz, chirality, and flavor structures. We study the representative operators with coefficients $C_{SV},C_{PV},C_{SA},C_{PA}$:
\begin{align}
  \cO_{\Gamma V}^{(9)}
  &=[\overline e_+\Gamma\mu]
  (\bar d\gamma_\alpha u)(\bar d\gamma^\alpha u),
  \nonumber\\
  \cO_{\Gamma A}^{(9)}
  &=[\overline e_+\Gamma\mu]
  (\bar d\gamma_\alpha\gamma_5u)
  (\bar d\gamma^\alpha\gamma_5u),
  \qquad \Gamma=1,\gamma_5,
  \label{eq:example-operators}
\end{align}
where $e_+$ stands for the positron-electron field by considering the positron as a particle with a bispinor wavefunction $u_+$.

\subsection{First-order transition element}

We calculate from scratch by starting with the definition of the first-order transition matrix element:
\begin{equation}
  S_{fi}^{(1)}
  =i\int\dd^4x\,
  \left\langle \Psi_f;e^+(\bp_e,s_e)\right|
  \cL_{\Delta L=2}^{(9)}(x)
  \left|\Psi_i;\mu^-_{1s}(s_\mu)\right\rangle .
  \label{eq:first-order-S}
\end{equation}
Here $\Psi_i$ is the unit-normalized initial nuclear wave function with the subscript $i$ denoting its complete quantum numbers: nuclear mass and charge numbers $(A,Z)$, total spin and magnetic number $(J_i,M_i)$, and mass $\mathsf M_i$. Similarly, the final nuclear wave function $\Psi_f$ has the quantum numbers $(A,Z-2),~(J_f,M_f)$, and energy $E_f$.
The bound $1s$ muon has spin $s_\mu$, mass $m_\mu$, and energy
\begin{equation}
  E_\mu=m_\mu-B_\mu ,
  \label{eq:bound-muon-energy}
\end{equation}
where $B_\mu>0$ is its atomic binding energy.
The emitted positron has an asymptotic four-momentum
$p_e^\alpha=(E_e,\bp_e)$, spin $s_e$, and mass $m_e$. Its magnitude and direction of momentum will be denoted as
$p_e=|\bp_e|$ and $\hat\bp_e=\bp_e/p_e$.

Let $H_0$ contain the nuclear Hamiltonian and static Coulomb fields that bind the muon and distort the positron, but not
$\cL_{\Delta L=2}^{(9)}$. Both initial and final states are energy eigenstates:
\begin{align}
  H_0|i\rangle&=(\mathsf M_i+E_\mu)|i\rangle,
  \nonumber\\
  H_0|f\rangle&=(E_f+E_e)|f\rangle .
  \label{eq:external-energies}
\end{align}
The time translation therefore gives
\begin{align}
  & \langle f|\cL_{\Delta L=2}^{(9)}(t,\bx)|i\rangle
  \nonumber\\[-2pt]
  ={}&e^{i(E_f+E_e-\mathsf M_i-E_\mu)t}
  \langle f|\cL_{\Delta L=2}^{(9)}(0,\bx)|i\rangle .
  \label{eq:time-translation}
\end{align}
The time integral in \cref{eq:first-order-S} can now be done
to yield the energy conservation $\delta$ function:
\begin{align}
  S_{fi}^{(1)}
  ={}&i2\pi\delta(\mathsf M_i+m_\mu-B_\mu-E_f-E_e)
  \cM_{fi}^{s_es_\mu},
  \label{eq:S-matrix}\\
  \cM_{fi}^{s_es_\mu}
  ={}&\int\dd^3x\,
  \langle f|\cL_{\Delta L=2}^{(9)}(0,\bx)|i\rangle ,
  \label{eq:reduced-amplitude}
\end{align}
where $\cM_{fi}^{s_es_\mu}$ is the reduced amplitude and energy conservation implies
\begin{equation}
  E_e=\mathsf M_i+m_\mu-B_\mu-E_f,
  \qquad
  p_e=\sqrt{E_e^2-m_e^2}.
  \label{eq:energy-conservation}
\end{equation}

\subsection{Exclusive rate}

With relativistically normalized positron states and unit-normalized nuclear
states, the exclusive rate is
\begin{align}
  \Gamma_{i\to f}
  ={}&\frac{p_e}{8\pi^2}\frac{1}{2J_i+1}
  \sum_{M_iM_f}\int\dd\Omega_e\,
  \frac12\sum_{s_\mu s_e}
  \left|\cM_{fi}^{s_es_\mu}(\hat\bp_e)\right|^2 .
  \label{eq:master-rate}
\end{align}
Here $\dd\Omega_e$ is the solid-angle element of the asymptotic direction $\hat\bp_e$. The nuclear magnetic substates and the muon spin are averaged for an unpolarized target with unresolved hyperfine structure, and the final nuclear magnetic substates and positron spin are summed. Experiments usually quote
\begin{equation}
  R_{\mu e^+}=\frac{\sum_f\Gamma_{i\to f}}{\Gamma_{\rm capt}},
  \label{eq:conversion-ratio}
\end{equation}
where the set of daughter states in the numerator must follow the experimental signal definition and
$\Gamma_{\rm capt}$ is the ordinary muon-capture rate.

\section{Spatial reduction and momentum flow}
\label{sec:leptons-routing}

At fixed time, the matrix element of each operator in
Eq.~\eqref{eq:example-operators} factorizes into a leptonic transition density and a hadronic one. For
$\Gamma=1,\gamma_5$, we denote
\begin{align}
  L_\Gamma^{s_es_\mu}(\bx)
  &={}\overline u_+(\bx,s_e)\Gamma u_\mu(\bx,s_\mu),
  \label{eq:leptonic-density}\\
  H_X^{fi}(\bx)
  &={}\langle\Psi_f|J_X^\alpha(0,\bx)J_{X,\alpha}(0,\bx)
  |\Psi_i\rangle,
  \label{eq:hadronic-density}
\end{align}
with $X=V,A$ for the vector and axial cases, respectively.
The corresponding part of the reduced amplitude is
\begin{equation}
  \cM_{fi}^{s_es_\mu}
  =\frac{G_F^2}{2m_p}\sum_{\Gamma,X}C_{\Gamma X}
  \int\dd^3x\,L_\Gamma^{s_es_\mu}(\bx)H_X^{fi}(\bx).
  \label{eq:spatial-factorization}
\end{equation}
We next reduce the lepton and hadron factors separately and then perform the interaction-coordinate integral.

\subsection{Lepton wave functions in the nuclear volume}

To get a sense of various scales, the characteristic radius of the bound muon may be estimated by a point-Coulomb potential as
\begin{equation}
  a_\mu=(Z\alpha m_{\mu A})^{-1}
  \simeq 256\ {\rm fm}/Z,
  \label{eq:muon-Bohr-radius}
\end{equation}
where $m_{\mu A}$ is the muon--nucleus reduced mass.  This gives approximately $20$ fm for aluminum and
$12$ fm for titanium, to be compared with their nuclear radii of about $3.6$ and $4.4$ fm. Thus, within the nuclear volume, the muon wave function varies slowly on both the typical nuclear-size scale and the positron de Broglie wavelength,
$1/p_e\simeq2$ fm for $p_e\simeq100$ MeV.  At the accuracy pursued here we neglect the muon momentum and replace the dominant $1s$ component by a nuclear average.
In the Dirac representation, we write
\begin{equation}
  u_\mu(\bx,s_\mu)\simeq
  \langle g_\mu\rangle_\rho
  \begin{pmatrix}\chi_{s_\mu}\\0
  \end{pmatrix},
  \quad
  \langle g_\mu\rangle_\rho=
  \frac{\int\dd^3x\,\rho(\bx)g_\mu(\bx)}
  {\int\dd^3x\,\rho(\bx)} .
  \label{eq:average-muon}
\end{equation}
The weight $\rho(\bx)$ is ultimately transition dependent.  Varying it among a charge, matter, and leading transition density provides a practical estimate of the averaging uncertainty~\cite{WeinbergFeinberg1959,Kosmas1994,Domin2004,
KitanoKoikeOkada2002,Haxton2023}.

The outgoing positron is more delicate because
$p_eR_A$ is of order unity. For example, at
$p_e=100$ MeV, the radii quoted above give
$p_eR_A\simeq1.8$ for aluminum and $2.2$ for titanium. An often used rate-level Fermi function, to be multiplied onto the rate,
\begin{align}
  F_+(Z_f,E_e)={}&
  \frac{2(1+\gamma_f)}{\Gamma(2\gamma_f+1)^2}
  (2p_eR_c)^{2(\gamma_f-1)}
  \nonumber\\[-2pt]
  &\times e^{-\pi y_f}
  \left|\Gamma(\gamma_f-iy_f)\right|^2,
  \label{eq:fermi-function}\\
  \gamma_f={}&\sqrt{1-(\alpha Z_f)^2},
  \qquad
  y_f=\frac{\alpha Z_fE_e}{p_e},
  \nonumber
\end{align}
with daughter charge $Z_f=Z-2$ and representative charge radius $R_c$, accounts only for an overall
$s_{1/2}$ Coulomb normalization~\cite{Doi1985,Domin2004,GeibMerleZuber2017}.
It does not incorporate spatially varying phase, partial-wave-dependent magnitudes and phases, finite-size radial variation, or redistribution among nuclear multipoles.

We instead advocate the effective-momentum approximation (EMA), an analytic intermediate step between a plane wave and an exact finite-size Dirac solution. Such approximations have been developed for lepton-nucleus scattering and tested in muon-electron conversion kinematics~\cite{Traini2001,WallaceTjon2008,Haxton2023,HaxtonRule2025}. For the repelled positron, define
\begin{equation}
  \eeff=E_e-\overline V_C,\quad
  \peff=\sqrt{\eeff^2-m_e^2},\quad
  N_e=\frac{\peff}{p_e},
  \label{eq:effective-momentum}
\end{equation}
where $\overline V_C>0$ is the repulsive Coulomb potential of the daughter nucleus averaged over the nuclear charge distribution. The effective and asymptotic momenta have the same direction,
\begin{equation}
  \bpeff=\peff\,\hat\bp_e .
  \label{eq:effective-vector}
\end{equation}
With EMA, the outgoing positron has the wavefunction:
\begin{align}
  & u_+(\bx,s_e)
  \simeq N_e\xi_e
  \begin{pmatrix}
    \chi_{s_e}\\[2pt]
    \eta_e\,\bsigma\cdot\widehat\bp_e\,\chi_{s_e}
  \end{pmatrix}
  e^{i\bpeff\cdot\bx},
  \label{eq:positron-bispinor}\\
  & \xi_e=\sqrt{E_e+m_e},
  \quad
  \eta_e=\frac{p_e}{E_e+m_e}.
  \nonumber
\end{align}
Thus, $N_e$ and the spatial phase contain the effective momentum.
For the ultrarelativistic positron considered here, $\xi_e\approx\sqrt{E_e}$ and $\eta_e\approx 1$.
In \cref{tab:positron-treatments} we summarize common treatments of the positron.

\begin{table*}[t]
  \caption{Physics retained in common treatments of the outgoing positron.}
  \label{tab:positron-treatments}
  \begin{ruledtabular}
    \begin{tabular}{lccc}
      Treatment & Coulomb magnitude & Spatial dependence & Nuclear multipoles\\
      \hline
      Plane wave & No & $e^{i\bp_e\cdot\bx}$ & $j_L(p_eR)$\\
      Fermi-factor rescaling & Overall $F_+^{1/2}$ from $s_{1/2}$
      & $e^{i\bp_e\cdot\bx}$ & $j_L(p_eR)$\\
      Effective-momentum approximation & $N_e=\peff/p_e$
      & $e^{i\bpeff\cdot\bx}$ & $j_L(\peff R)$\\
      Exact Dirac wave & Partial-wave dependent
      & Coulomb radial waves and phases & Full Coulomb multipoles\\
    \end{tabular}
  \end{ruledtabular}
\end{table*}

The scalar and pseudoscalar densities now separate as
\begin{subequations}
  \begin{align}
    &\overline u_+u_\mu
    \simeq\mathcal N_\ell e^{-i\bpeff\cdot\bx}\ell_S,
    \quad\ell_S=\chi_{s_e}^\dagger\chi_{s_\mu},
    \label{eq:lepton-factors-S}
    \\
    &\overline u_+\gamma_5u_\mu
    \simeq\mathcal N_\ell e^{-i\bpeff\cdot\bx}\ell_P,
    \quad\ell_P=-\eta_e\chi_{s_e}^\dagger
    \bsigma\cdot\widehat\bp_e\chi_{s_\mu},
    \label{eq:lepton-factors-P}
  \end{align}
\end{subequations}
where $\mathcal N_\ell=N_e\xi_e\langle g_\mu\rangle_\rho$.
Their unpolarized spin sums and averages are
\begin{align}
  \frac12\sum_{s_es_\mu}|\ell_S|^2=1,
  &\qquad
  \frac12\sum_{s_es_\mu}|\ell_P|^2=\eta_e^2,
  \nonumber\\[-2pt]
  \frac12\sum_{s_es_\mu}\ell_S\ell_P^*=0.
  \label{eq:lepton-spin-sums}
\end{align}

\subsection{Impulse approximation and coordinate-space currents}

We next replace each quark current by a sum of one-body nucleon currents.  This impulse approximation resolves the local product of quark currents into operators acting on an ordered pair of bound nucleons.  For an on-shell $p\to n$ transition, let $q^\alpha=p'^\alpha-p^\alpha$ and $t=q^2$. The complete charged currents are
\begin{widetext}
  \begin{subequations}
    \begin{align}
      \langle n(p')|J_V^\alpha|p(p)\rangle
      &={}
      \bar u_n(p')\tau^-
      \left[g_V(t)\gamma^\alpha
        +ig_M(t)\frac{\sigma^{\alpha\nu}q_\nu}{2m_N}
      +g_S(t)\frac{q^\alpha}{m_N}\right]u_p(p),
      \label{eq:general-vector-current}\\
      \langle n(p')|J_A^\alpha|p(p)\rangle
      &={}
      \bar u_n(p')\tau^-
      \left[g_A(t)\gamma^\alpha
        +g_P(t)\frac{q^\alpha}{2m_N}
      +ig_T(t)\frac{\sigma^{\alpha\nu}q_\nu}{2m_N}\right]
      \gamma_5u_p(p).
      \label{eq:general-axial-current}
    \end{align}
    \label{eq:general-currents}
  \end{subequations}
\end{widetext}
Here $g_V,~g_M,~g_S,~g_A,~g_P,~g_T$ are respectively the vector, weak-magnetism, induced-scalar, axial, induced-pseudoscalar, and induced-tensor form factors. The isospin operator $\tau^-$ converts a proton into a neutron.

The relation between this momentum-space matrix element and an operator acting on bound nucleons is most transparent after a three-dimensional Fourier transform. In the instantaneous approximation $q_i^0=0$, we define the Pauli operator
$\mathcal J_{X,a}^\alpha(\bq_i,\bK_a)$ by reducing
Eq.~\eqref{eq:general-currents}, including all form factors at $Q_i^2=|\bq_i|^2$, where the subscript
$i=1,2$ will be associated with the two hadronic currents from dim-9 operators. The corresponding current density is
\begin{equation}
  J_X^\alpha(0,\bx)\rightarrow
  \sum_a\tau_a^-\int\frac{\dd^3q_i}{(2\pi)^3}
  e^{-i\bq_i\cdot(\bx-\br_a)}
  \mathcal J_{X,a}^\alpha(\bq_i,\bK_a).
  \label{eq:coordinate-current}
\end{equation}
Here, the sum runs over the nucleons bound in the nucleus, $\br_a$ is the coordinate of nucleon $a$, $\bx$ is the coordinate of the effective local-interaction point, and $\bq_i$ is the momentum pointing from $\bx$ to
$\br_a$. The Fourier phase follows from
$\langle p'|J(0,\bx)|p\rangle\propto
e^{-i(\bp'-\bp)\cdot\bx}$ at fixed time. The coordinate $\br_a$ and the spin and isospin operators act on the nucleon $a$ within the nuclear wave function. If a form factor is constant and the corresponding current component contains no explicit $\bq_i$, its Fourier integral is $\delta^{(3)}(\bx-\br_a)$: the current is then localized at a point nucleon. Momentum-dependent form factors smear this delta function over the nucleon size, while explicit powers of $\bq_i$ generate derivatives of form factors defined in coordinate space. The average nucleon momentum
\begin{equation}
  \bK_a=\bp_a'+\bp_a
  \rightarrow-i\overleftrightarrow{\boldsymbol\nabla}_a
  \equiv-i(\overrightarrow{\boldsymbol\nabla}_a-
  \overleftarrow{\boldsymbol\nabla}_a)
  \label{eq:nucleon-average-momentum}
\end{equation}
is instead a two-sided derivative acting on nuclear wave functions.

The product of two currents at the same interaction coordinate is therefore
\begin{equation}
  \begin{aligned}
    &J_X^\alpha(0,\bx)J_{X,\alpha}(0,\bx)
    \\[-2pt]
    \rightarrow
    &\sum_{a\ne b}\tau_a^-\tau_b^-
    \int\frac{\dd^3q_1\dd^3q_2}{(2\pi)^6}
    \\[-2pt]
    &\times e^{-i\bq_1\cdot(\bx-\br_a)}
    e^{-i\bq_2\cdot(\bx-\br_b)}
    \mathcal K_{X,ab}(\bq_1,\bq_2),
  \end{aligned}
  \label{eq:two-current-density}
\end{equation}
where
$\mathcal K_{X,ab}=g_{\alpha\beta}\mathcal J_{X,a}^\alpha(\bq_1,\bK_a)
\mathcal J_{X,b}^\beta(\bq_2,\bK_b)$. The restriction
$a\ne b$ is automatic for isospin-$1/2$ nucleons because $(\tau_a^-)^2=0$. Most importantly, each current retains its own momentum and its own form-factor argument.

\subsection{The interaction-coordinate integral}

Inserting Eqs.~\eqref{eq:lepton-factors-S}, \eqref{eq:lepton-factors-P}, and \eqref{eq:two-current-density} into Eq.~\eqref{eq:spatial-factorization}
makes the momentum constraint explicit:
\begin{align}
  \int\dd^3x\,
  e^{-i\bpeff\cdot\bx}
  e^{-i(\bq_1+\bq_2)\cdot\bx}
  ={}&(2\pi)^3
  \delta^{(3)}(\bpeff+\bq_1+\bq_2).
  \label{eq:spatial-delta}
\end{align}
Thus, locality fixes only the sum
\begin{equation}
  \bq_1+\bq_2=-\bpeff,
  \label{eq:momentum-sum}
\end{equation}
but does not require equal momentum transfers at the two nucleons. We introduce one unconstrained sharing momentum
$\bq$ by
\begin{align}
  \bq_1=\bq-\frac{\bpeff}{2},
  &\qquad
  \bq_2=-\bq-\frac{\bpeff}{2},
  \label{eq:momentum-routing}
\end{align}
so that
\begin{equation}
  Q_1^2=\left|\bq-\frac{\bpeff}{2}\right|^2,
  \qquad
  Q_2^2=\left|\bq+\frac{\bpeff}{2}\right|^2.
  \label{eq:two-Q}
\end{equation}
The remaining nucleon-coordinate phase becomes
\begin{equation}
  e^{i\bq_1\cdot\br_a+i\bq_2\cdot\br_b}
  =e^{-i\bpeff\cdot\bR_{ab}}e^{i\bq\cdot\br_{ab}},
  \label{eq:two-current-phase}
\end{equation}
where
\begin{equation}
  \bR_{ab}=\frac{\br_a+\br_b}{2},
  \qquad
  \br_{ab}=\br_a-\br_b ,
  \label{eq:pair-coordinates}
\end{equation}
are the center-of-mass coordinate and the relative coordinate of the nucleon pair $ab$ associated with the momenta $\bpeff$ and $\bq$ respectively. After the interaction-coordinate integral, a generic two-current contribution to the nuclear operator consequently has the form
\begin{align}
  \widehat\cO_\Gamma(\bpeff)
  =&\sum_{a\ne b}\tau_a^-\tau_b^-
  e^{-i\bpeff\cdot\bR_{ab}}
  \int\frac{\dd^3q}{(2\pi)^3}e^{i\bq\cdot\br_{ab}}
  \nonumber\\[-2pt]
  &\times\left[
    C_{\Gamma V}\mathcal K_{V,ab}(\bq_1,\bq_2)
    +C_{\Gamma A}\mathcal K_{A,ab}(\bq_1,\bq_2)
  \right].
  \label{eq:generic-two-body-operator}
\end{align}
This derivation explains both the single $\bq$ integral and the two distinct form-factor momenta.  Replacing $Q_1^2$ and $Q_2^2$ by a common argument is an additional approximation, not a consequence of the locality of the dim-9 interaction.

\section{First-class pion-pole-improved currents}
\label{sec:PI-currents}

\subsection{Working first-class reduction}

We now reduce the complete nucleon currents in
Eq.~\eqref{eq:general-currents} to the Pauli operators needed in nuclear wave
functions.  This step determines the kernels
$\mathcal K_{X,ab}$ left general in Eq.~\eqref{eq:generic-two-body-operator}.
The induced scalar $g_S$ and tensor $g_T$ are second-class form factors and vanish in the isospin limit. The first-class set is $g_V,g_M,g_A,g_P$.  In the convention of
Eq.~\eqref{eq:general-currents}, $g_V(0)=1$,
$g_M(0)=\kappa_p-\kappa_n\simeq3.706$, and $g_A(0)\simeq1.27$. For spacelike
transfer, $Q^2=-t>0$, the leading PCAC/pion-pole relation reads~\cite{BernardElouadrhiriMeissner2002}
\begin{equation}
  g_P(-Q^2)\simeq\frac{4m_N^2g_A(-Q^2)}{Q^2+m_\pi^2}.
  \label{eq:pion-pole-form-factor}
\end{equation}
We label the two hadronic-current insertions by $i=1,2$ and abbreviate
\begin{align}
  V_i&=g_V(-Q_i^2),&M_i&=g_M(-Q_i^2),
  \nonumber\\[-2pt]
  A_i&=g_A(-Q_i^2),&P_i&=g_P(-Q_i^2),
  \label{eq:form-factor-shorthand}
\end{align}
and impose the first-class limit
\begin{equation}
  g_S(-Q_i^2)=g_T(-Q_i^2)=0,
  \label{eq:first-class-limit}
\end{equation}
at the one-current level. In instantaneous reduction, with $q_i^0=0$ and relativistic external-state normalization removed, the retained Pauli operators are
\begin{subequations}
  \begin{align}
    \mathcal V_i^0&=V_i,
    &\boldsymbol{\mathcal V}_i&=\bB_i+i\bW_i,
    \label{eq:PI-vector-current}\\
    \bB_i&=V_i\frac{\bK_i}{2m_N},
    &\bW_i&=(V_i+M_i)\frac{\bsigma_i\times\bq_i}{2m_N},
    \label{eq:PI-vector-components}\\
    \mathcal A_i^0&=a_i^0=A_i\frac{\bsigma_i\cdot\bK_i}{2m_N},
    &\boldsymbol{\mathcal A}_i&=A_i\bsigma_i-\bL_i,
    \label{eq:PI-axial-current}\\
    \bL_i&=P_i\frac{\bq_i(\bq_i\cdot\bsigma_i)}{4m_N^2},
    \label{eq:PI-longitudinal-current}
  \end{align}
  \label{eq:PI-one-body-currents}
\end{subequations}
\par\noindent
where $\bB_i$ is the convection current, $\bW_i$ the magnetization current, $a_i^0$ the ordinary axial charge, and $\bL_i$ the induced-pseudoscalar spin-longitudinal current. The current-slot label means
$\bsigma_{1,2}=\bsigma_{a,b}$ and $\bK_{1,2}=\bK_{a,b}$.
The complete six-form-factor version of this reduction is given in Appendix~\ref{app:general-currents}.

\subsection{Two-current kernels and power counting}

Within the PI prescription, the Lorentz-contracted first-class
kernels are
\begin{subequations}
  \begin{align}
    \mathcal K_{V,ab}^{\rm PI}&=V_1V_2-\bB_1\cdot\bB_2+\bW_1\cdot\bW_2
    \nonumber
    \\
    &\quad-i(\bB_1\cdot\bW_2+\bW_1\cdot\bB_2),
    \label{eq:PI-vector-kernel}
    \\
    \mathcal K_{A,ab}^{\rm PI}
    &=a_1^0a_2^0-A_1A_2\bsigma_a\cdot\bsigma_b
    \nonumber
    \\
    &+A_1\bsigma_a\cdot\bL_2+A_2\bL_1\cdot\bsigma_b
    -\bL_1\cdot\bL_2.
    \label{eq:PI-axial-kernel}
  \end{align}
\end{subequations}
For notational brevity, in this section and the next we denote simply $\bp=\bpeff$, $p=\peff$, and $\widehat\bp=\widehat\bp_e$. The ordered-pair nuclear operator for either lepton bilinear is
\begin{align}
  \widehat\cO_{\Gamma}^{\rm PI}(\bp)
  ={}&\sum_{a\ne b}\tau_a^-\tau_b^-
  e^{-i\bp\cdot\bR_{ab}}
  \int\frac{\dd^3q}{(2\pi)^3}e^{i\bq\cdot\br_{ab}}
  \nonumber\\[-2pt]
  &\times\left[C_{\Gamma V}\mathcal K_{V,ab}^{\rm PI}
  +C_{\Gamma A}\mathcal K_{A,ab}^{\rm PI}\right].
  \label{eq:PI-two-body-operator}
\end{align}
Every form factor in Eqs.~\eqref{eq:PI-vector-kernel} and
\eqref{eq:PI-axial-kernel} is evaluated at its own $Q_i^2$.

For later implementation, it is useful to expose the eight primitive PI channels.  With $G_i=V_i+M_i$, the terms in the brackets of \cref{eq:PI-two-body-operator} are
\begin{widetext}
  \begin{subequations}
    \begin{align}
      \mathcal P_{VV}={}&C_{\Gamma V}V_1V_2,
      \label{eq:channel-VV}\\
      \mathcal P_{BB}={}&-\frac{C_{\Gamma V}V_1V_2}{4m_N^2}
      \bK_a\cdot\bK_b,
      \label{eq:channel-BB}\\
      \mathcal P_{WW}={}&\frac{C_{\Gamma V}G_1G_2}{4m_N^2}
      \left[(\bsigma_a\cdot\bsigma_b)(\bq_1\cdot\bq_2)
      -(\bsigma_a\cdot\bq_2)(\bsigma_b\cdot\bq_1)\right],
      \label{eq:channel-WW}\\
      \mathcal P_{BW}={}&-\frac{iC_{\Gamma V}}{4m_N^2}
      \left[V_1G_2\bK_a\cdot(\bsigma_b\times\bq_2)
      +G_1V_2(\bsigma_a\times\bq_1)\cdot\bK_b\right],
      \label{eq:channel-BW}\\
      \mathcal P_{AA}={}&-C_{\Gamma A}A_1A_2\bsigma_a\cdot\bsigma_b,
      \label{eq:channel-AA}\\
      \mathcal P_{A0A0}={}&\frac{C_{\Gamma A}A_1A_2}{4m_N^2}
      (\bsigma_a\cdot\bK_a)(\bsigma_b\cdot\bK_b),
      \label{eq:channel-A0A0}\\
      \mathcal P_{AP}={}&\frac{C_{\Gamma A}}{4m_N^2}
      \left[A_1P_2(\bsigma_a\cdot\bq_2)(\bsigma_b\cdot\bq_2)
      +P_1A_2(\bsigma_a\cdot\bq_1)(\bsigma_b\cdot\bq_1)\right],
      \label{eq:channel-AP}\\
      \mathcal P_{PP}={}&-\frac{C_{\Gamma A}P_1P_2}{16m_N^4}
      (\bq_1\cdot\bq_2)
      (\bsigma_a\cdot\bq_1)(\bsigma_b\cdot\bq_2).
      \label{eq:channel-PP}
    \end{align}
    \label{eq:PI-channels}
  \end{subequations}
\end{widetext}
This list is closed under pair exchange and involves only six ordered form-factor families:
\begin{equation}
  \nonumber
  (V,V),\quad(V,G),\quad(G,G),\quad(A,A),\quad(A,P),\quad(P,P).
\end{equation}

In a strict explicit inverse-mass count, $\bB_i$, $\bW_i$, and $a_i^0$ are $O(m_N^{-1})$, while $\bL_i$ appears to be $O(m_N^{-2})$. Equations \eqref{eq:PI-vector-kernel} and \eqref{eq:PI-axial-kernel} retain all products through the displayed non-pole order $m_N^{-2}$ and additionally retain $-\bL_1\cdot\bL_2$. The latter is nominally $m_N^{-4}$, but the pion pole in $P_i$ cancels the displayed $m_N^2$ in each one-current ratio,
\begin{align}
  \frac{|\bL_i|}{|A_i\bsigma_{i,L}|}
  \simeq\frac{Q_i^2}{Q_i^2+m_\pi^2}.
  \label{eq:pion-pole-ratio}
\end{align}
The $AP$ and $PP$ pieces are therefore successive terms in the product of two spin-longitudinal currents and should be kept together. This is the PI prescription. Note that this does not mean that \cref{eq:PI-one-body-currents} are a complete $O(m_N^{-2})$ Foldy--Wouthuysen expansion. The Darwin,
spin--orbit, external-state-normalization, and non-pole axial recoil terms at that order have not all been included. Neither have the many-body currents required by a specified chiral EFT. This will be briefly discussed in \cref{app:checks}.

\section{Exact multipoles for a $0^+\to 0^+$ transition}
\label{sec:zero-plus}

\subsection{Unequal-form-factor harmonics}
The unequal arguments in Eq.~\eqref{eq:two-Q} give nonpolynomial angular dependence. For an ordered form-factor product, we have
\begin{align}
  X(Q_1^2)Y(Q_2^2)
  &=\sum_{n=0}^\infty(2n+1)\mathcal F_n^{XY}(p,q)P_n(z),
  \label{eq:mixed-FF-expansion}
\end{align}
where $z=\hat\bp\cdot\hat\bq$ and
\begin{align}
  \mathcal F_n^{XY}(p,q)
  &=\frac12\int_{-1}^{1}\dd z\,P_n(z)
  X\!\left(q^2+\frac{p^2}{4}-pqz\right)
  \nonumber\\[-2pt]
  &\quad\times Y\!\left(q^2+\frac{p^2}{4}+pqz\right).
  \label{eq:mixed-FF-coefficient}
\end{align}
Interchanging the two hadronic-current labels gives
\begin{equation}
  \mathcal F_n^{YX}(p,q)=(-1)^n\mathcal F_n^{XY}(p,q).
  \label{eq:mixed-FF-exchange}
\end{equation}
Thus equal products contain only even $n$, whereas ordered $VG$ and $AP$ products may contain odd $n$ before pair exchange.  This is the generalization of the even relative-coordinate multipoles found for the leading equal-form-factor kernels.

For convenience, we use Racah-normalized harmonics
\begin{equation}
  C_{LM}(\widehat{\mathbf n})=
  \frac{\sqrt{4\pi}}{\hat L} 
  Y_{LM}(\widehat{\mathbf n}),
  \label{eq:Racah-harmonic}
\end{equation}
with $\hat x=\sqrt{2x+1}$.
After the explicit momenta and spins in each channel of
Eq.~\eqref{eq:PI-channels} have been decomposed into irreducible tensors, one may write a primitive scalar as
\begin{equation}
  \mathcal P_\rho=w_\rho(p,q)
  \left[[C_{l_p}(\widehat\bp)\otimes C_{l_q}(\widehat\bq)]_S
  \otimes\Xi_{\rho S}(a,b)\right]_{00},
  \label{eq:primitive-channel}
\end{equation}
where $\Xi_{\rho S}$ denotes an irreducible nuclear spin–derivative tensor of rank $S$. Here $\rho$ actually depends on a set of numbers such as $l_p,~l_q,~S$ and a product of form factors in a primitive term of a channel.
But for notational brevity, we will only indicate the channel explicitly,
$\rho\in\mathcal C_{\rm PI}$, with
\begin{equation}
  \mathcal C_{\rm PI}=\{VV,BB,WW,BW,AA,A0A0,AP,PP\},
  \label{eq:PI-channel-set}
\end{equation}
and the product of form factors $X(Q_1^2)Y(Q_2^2)$ in each primitive term of \cref{eq:PI-channels} is not yet included in \cref{eq:primitive-channel}. The fully expanded primitive dictionary for the eight PI channels is given in \cref{tab:PI-primitive-dictionary} of \cref{app:multipoles}.

\subsection{Scalar nuclear operator}

For arbitrary nuclear states, we expand the fixed-direction operator as
\begin{align}
  \widehat\cO_\Gamma^{\rm PI}(\bp)
  =\sum_{JM}C^*_{JM}(\widehat\bp)
  \widetilde\cO_{\Gamma;JM}^{\rm PI}(p).
  \label{eq:PI-multipole-expansion}
\end{align}
Here
$C^*_{JM}(\widehat\bp)=(-1)^M C_{J,-M}(\widehat\bp)$; this standard convention ensures that
$\widetilde\cO_{\Gamma;JM}^{\rm PI}$ transforms as the $M$ component of a rank-$J$ nuclear tensor.
The parity-even Lorentz contractions considered here produce a nuclear tensor of parity $(-1)^J$, so
\begin{align}
  |J_i-J|\le J_f\le J_i+J,
  \qquad \pi_f=\pi_i(-1)^J.
  \label{eq:selection-rules}
\end{align}
For $0_i^+\to0_f^+$, only $J=M=0$ survives. The two external Clebsch--Gordan coefficients become
\begin{align}
  \braket{L0\,L_p0|00}&=\frac{(-1)^L}{\hat L}\delta_{L_pL},
  \nonumber\\
  \braket{Lm\,L_pm_p|00}&=\frac{(-1)^{L-m}}{\hat L}
  \delta_{L_pL}\delta_{m_p,-m}.
  \label{eq:zero-CG}
\end{align}
The subscript $00$ in $[\cdots]_{00}$ of \cref{eq:primitive-channel} and below denotes a spherical tensor of rank zero and projection zero. All external magnetic sums can therefore be performed and the factor $2L+1$ in the general expression cancels. The exact scalar PI operator is
\begin{align}
  \nonumber
  &\widetilde\cO_{\Gamma;00}^{\rm PI}(p)
  \\
  \nonumber
  =&\sum_{a\ne b}\tau_a^-\tau_b^-
  \sum^{\rho\in\mathcal C_{\rm PI}}_{L,L_q,S}
  (-i)^Lj_L(pR_{ab})
  \mathcal H^{\rho,\rm PI}_{L L_qS}(r_{ab};p)
  \\
  &\times\left[
    [C_L(\widehat\bR_{ab})\otimes C_{L_q}(\widehat\br_{ab})]_S
    \otimes\Xi_{\rho S}(a,b)
  \right]_{00},
  \label{eq:PI-zero-plus-operator}
\end{align}
where the Fourier-Bessel kernel $\mathcal H^{\rho,\rm PI}_{L_pL_qS}(r;p)$ is derived in \cref{app:multipoles}. \cref{eq:PI-zero-plus-operator} is the principal nuclear result of this work. Relative to the arbitrary-transition result, it contains no $J,M,L_p$, or magnetic-index sums.  It does, however, retain $L>0$, $L_q>0$, and $S>0$ internal tensors.  Finite positron momentum therefore does not justify an $s$-wave or monopole-only nuclear truncation even for a $0^+\to0^+$ transition.

At $p=0$, $j_L(0)=\delta_{L0}$, and \cref{eq:PI-zero-plus-operator} degenerates to
\begin{align}
  \widetilde\cO_{\Gamma;00}^{\rm PI}(0)
  =&\sum_{a\ne b}\tau_a^-\tau_b^-
  \sum^{\rho\in\mathcal C_{\rm PI}}_S
  \mathcal H^{\rho,\rm PI}_{0SS}(r_{ab};0)
  \nonumber\\
  &\qquad\times
  [C_S(\widehat\br_{ab})\otimes\Xi_{\rho S}(a,b)]_{00}.
  \label{eq:PI-zero-plus-pzero}
\end{align}
Only the leading $VV$ and $AA$ scalar channels additionally force $S=0$; explicit momentum and derivative tensors may retain $S>0$.

For comparison, if only the leading vector charge or axial spin is retained, the finite-size scalar kernel is
\begin{align}
  \nonumber
  H_X(r,p,z_r)&=\int\frac{\dd^3q}{(2\pi)^3}e^{i\bq\cdot\br}
  F_X(Q_1^2)F_X(Q_2^2),
  \\
  &=\sum_{\lambda\ \mathrm{even}}(2\lambda+1)
  h_{X\lambda}(r,p)P_\lambda(z_r),
  \label{eq:leading-kernel}
\end{align}
with $z_r=\hat\br\cdot\hat\bp$. The PI construction promotes this leading Fermi/Gamow--Teller pair to the eight-channel basis in \cref{eq:PI-channel-set}; yet it does not alter the logic of the independent-momentum expansion.

\subsection{Scalar two-body transition density and rate}

Let $c^\dagger_{at}$ be the spherical one-nucleon creation tensor and define the normalized pair-creation and pair-annihilation tensors by
\begin{align}
  A^\dagger_{JM}(ab;t)&=
  \frac{[c^\dagger_{at}\otimes c^\dagger_{bt}]_{JM}}
  {\sqrt{1+\delta_{ab}}},
  \nonumber\\[-2pt]
  \widetilde A_{JM}(ab;t)&=(-1)^{J-M}A_{J,-M}(ab;t).
  \label{eq:pair-operators}
\end{align}
Here, $A_{JM}$ annihilates the normalized pair, while the tilde denotes its spherical-tensor conjugate, with the phase chosen so that $\widetilde A_{JM}$ transforms as a rank-$J$ tensor, and $\delta_{ab}$ supplies the identical-orbit normalization.
In this subsection only, $a,b,c,d$ label spherical single-particle  orbits rather than the first-quantized nucleons used above; each orbit label includes its radial and angular quantum numbers, and $t=n,p$ labels neutron or proton isospin. For a scalar transition, we denote the charge-changing two-body
transition density (TBTD) at pair angular momentum $J_c$ by
\begin{align}
  & \mathrm{TBTD}_{fi}^{(0)}(ab,cd;J_c)
  \nonumber\\[-2pt]
  =&
  \left\langle0_f^+\right\|
  [A^\dagger_{J_c}(ab;n)\otimes
  \widetilde A_{J_c}(cd;p)]_0
  \left\|0_i^+\right\rangle .
  \label{eq:scalar-TBTD}
\end{align}
The subscript $0$ on the square bracket is its tensor rank, and the
double bars, as usual, denote a reduced matrix element. The above operator creates a neutron pair in the daughter, while annihilating a proton pair in the parent.  We denote by $\widetilde\cO_{\Gamma}^{\rm PI,(0)}(p)$ the rank-zero many-body tensor whose only spherical component is
$\widetilde\cO_{\Gamma;00}^{\rm PI}(p)$ in
Eq.~\eqref{eq:PI-zero-plus-operator}, and by
$\widetilde o_{\Gamma}^{\rm PI,(0)}(p)$ its two-nucleon counterpart.
With normalized antisymmetrized (AS) pair states, their contraction is
\begin{align}
  & \left\langle0_f^+\right\|
  \widetilde\cO_{\Gamma}^{\rm PI,(0)}(p)
  \left\|0_i^+\right\rangle
  \nonumber\\[-2pt]
  =&\sum_{\substack{a\le b,\,c\le d\\J_c}}
  \mathrm{TBTD}_{fi}^{(0)}(ab,cd;J_c)
  \nonumber\\[-2pt]
  &\times
  \left\langle ab;J_c\right\|
  \widetilde o_{\Gamma}^{\rm PI,(0)}(p)
  \left\|cd;J_c\right\rangle_{\rm AS},
  \label{eq:scalar-TBTD-handoff}
\end{align}
where the proton orbits $cd$ are annihilated in the parent and the neutron orbits $ab$ are created in the daughter. The same ordering and normalization must be used in the TBTD and the antisymmetrized two-body matrix element.

Denoting
\begin{equation}
  \mathcal M_\Gamma^{\rm PI}(p)=
  \left\langle0_f^+\right|
  \widetilde\cO_{\Gamma;00}^{\rm PI}(p)
  \left|0_i^+\right\rangle,
  \qquad \Gamma=S,P,
  \label{eq:scalar-NME}
\end{equation}
the reduced amplitude becomes direction independent,
\begin{equation}
  \cM_{fi}^{s_es_\mu}=
  \frac{G_F^2}{2m_p}\mathcal N_\ell
  \left[\ell_S\mathcal M_S^{\rm PI}
  +\ell_P\mathcal M_P^{\rm PI}\right],
  \label{eq:zero-plus-amplitude}
\end{equation}
and Eqs.~\eqref{eq:master-rate} and \eqref{eq:lepton-spin-sums} give
\begin{equation}
  \Gamma_{0^+\to0^+}=
  \frac{p_e}{2\pi}\left(\frac{G_F^2}{2m_p}\right)^2
  \mathcal N_\ell^2
  \left(|\mathcal M_S^{\rm PI}|^2
  +\eta_e^2|\mathcal M_P^{\rm PI}|^2\right).
  \label{eq:zero-plus-rate}
\end{equation}
There is no scalar--pseudoscalar interference after the unpolarized lepton spin sum. Vector and axial contributions within a fixed $\Gamma$ can still interfere through the Wilson coefficients and nuclear matrix elements.

\section{Phenomenological hierarchy and implementation}
\label{sec:implementation}

Now we compare the PI basis we employed in previous sections with some approximations.
The leading calculation retains only $VV$ and $AA$.  A strict selected-current $m_N^{-2}$ calculation adds $BB$, $WW$, $BW$, $A0A0$, and $AP$ but drops $PP$.  The preferred PI result restores $PP$ so that the spin-longitudinal
one-current structure is not broken. At a representative $Q=100~\mathrm{MeV}$, we estimate that
\begin{align}
  \frac{|\bW|}{|V|}\sim(1+M/V)\frac{Q}{2m_N}\simeq0.25,
  \nonumber\\[-2pt]
  \frac{|\bL|}{|A\bsigma_L|}\simeq
  \frac{Q^2}{Q^2+m_\pi^2}\simeq0.34,
  \label{eq:current-size-estimate}
\end{align}
using $m_N=939~\mathrm{MeV}$ and $M(0)=3.706$. These are coefficient-level estimates, not NME corrections, but they show why weak magnetism and the induced pseudoscalar cannot be discarded merely by counting the displayed inverse powers of $m_N$.  Their significant impact in $0\nu\beta\beta$ calculations provides a useful warning~\cite{SimkovicPantis1999,Graf2018}, although it does not determine their size in $\muep$ conversion.

Equation~\eqref{eq:mixed-FF-coefficient} shows that the current arguments contain both $\bpeff$ and the sharing momentum $\bq$. In the nuclear matrix element $\peff\sim100$ MeV, while the dominant $q\sim1/r_{ab}$ is also typically of order $100$--$200$ MeV. Thus, neglecting $\bpeff$ in either current argument, which would replace both $Q_i^2$ by $q^2$, is not justified. Furthermore, as a separate diagnostic of the angular dependence, setting $\peff=0$ in
Eq.~\eqref{eq:mixed-FF-coefficient} gives
\begin{equation}
  \mathcal F_n^{XY}(0,q)=\delta_{n0}X(q^2)Y(q^2).
  \label{eq:pzero-FF}
\end{equation}
This special zero-$\peff$ limit eliminates every induced relative multipole beyond the monopole for equal leading form factors. A numerical implementation should therefore converge the Legendre index $n$ and the orbital sums $L,L_q,S$ rather than set them to zero.

The natural first application is
$^{48}\mathrm{Ti}(0^+_{\rm gs})\to{}^{48}\mathrm{Ca}(0^+_{\rm gs})$.
The nuclear structure supplies the scalar charge-changing TBTDs and antisymmetrized two-body matrix elements in Eq.~\eqref{eq:scalar-TBTD-handoff}.
A transparent calculation should produce channel-by-channel results for $VV$, $AA$, weak magnetism, $AP$, and $PP$; compare leading, selected $m_N^{-2}$, and PI truncations; and document convergence in every angular cutoff. Until those TBTDs are supplied, the present work is an operator and nuclear-reduction framework rather than a conversion-rate prediction.

\section{Summary}
\label{sec:summary}

We have constructed a nuclear-level formulation of local dim-9 $\muep$ conversion that retains the independent momentum transfer at each nucleon current. The full covariant currents contain six form factors. The main calculation sets the second-class induced-scalar and induced-tensor currents to zero and uses a pion-pole-improved first-class reduction. This keeps convection, weak magnetism, axial charge, and the complete
spin-longitudinal pion-pole current generated by the displayed one-body operators.

The unequal form-factor arguments produce angular harmonics and nuclear operators that are identically absent in a common-momentum treatment.  An effective positron plane wave retains both the repulsive Coulomb normalization
and the shifted spatial phase, so its momentum enters every nuclear multipole. For $0^+\to0^+$ conversion the exact external angular reduction collapses to the scalar operator in Eq.~\eqref{eq:PI-zero-plus-operator}. This removes the
external rank and magnetic sums but leaves finite-momentum orbital, spin, and derivative tensors that must be converged in an actual calculation. The operator is ready for contraction with scalar charge-changing TBTDs.
We particularly encourage nuclear-structure groups to calculate the charge-changing TBTDs and antisymmetrized two-body matrix elements identified here, since those results would turn the present operator framework into quantitative conversion rates and would provide the most direct test of the proposed truncation hierarchy.

This short-range analysis is only the first part of our program. We will also investigate conversion driven by the standard light-Majorana-neutrino mass mechanism.  We expect that calculation to be substantially more difficult because the nonlocal neutrino propagator and its energy dependence must be combined consistently with the bound and continuum lepton wave functions and with nuclear intermediate-state dynamics.

\appendix

\section{Complete six-form-factor one-nucleon currents}
\label{app:general-currents}

For the current slot $i$, we define in addition to Eq.~\eqref{eq:form-factor-shorthand}
\begin{equation}
  S_i=g_S(-Q_i^2),\qquad T_i=g_T(-Q_i^2).
\end{equation}
Keeping the first nonzero static Pauli operator produced by every form factor gives
\begin{subequations}
  \begin{align}
    \nonumber
    \mathcal V_i^0&=V_i,
    \\
    \nonumber
    \boldsymbol{\mathcal V}_i&=\bB_i^{\rm full}+i\bW_i,
    \\
    \nonumber
    \bB_i^{\rm full}&=V_i\frac{\bK_i}{2m_N}
    +S_i\frac{\bq_i}{m_N},
    \\
    \bW_i&=(V_i+M_i)\frac{\bsigma_i\times\bq_i}{2m_N},
    \\[1em]
    \mathcal A_i^0&=a_i^{0,\rm full}
    =\frac{A_i\bsigma_i\cdot\bK_i
    +T_i\bsigma_i\cdot\bq_i}{2m_N},
    \nonumber\\
    \boldsymbol{\mathcal A}_i&=A_i\bsigma_i-\bL_i,
    \nonumber\\
    \bL_i&=P_i\frac{\bq_i(\bq_i\cdot\bsigma_i)}{4m_N^2}.
  \end{align}
  \label{eq:full-Pauli-currents}
\end{subequations}
\par\noindent
Terms proportional to $q_i^0$ are absent in the instantaneous realization. An energy-dependent reduction can restore them without changing the angular construction below. \cref{eq:full-Pauli-currents} retain the
leading occurrence of all six form factors, but not every non-pole term at a fixed order in $1/m_N$.

The complete vector kernel is
\begin{align}
  \mathcal K_{V,ab}^{\rm full}
  ={}&V_1V_2-\bB_1^{\rm full}\cdot\bB_2^{\rm full}
  +\bW_1\cdot\bW_2
  \nonumber\\[-2pt]
  &-i(\bB_1^{\rm full}\cdot\bW_2
  +\bW_1\cdot\bB_2^{\rm full}),
  \label{eq:full-vector-kernel}
\end{align}
while the axial kernel is
\begin{align}
  \mathcal K_{A,ab}^{\rm full}
  ={}&a_1^{0,\rm full}a_2^{0,\rm full}
  -A_1A_2\bsigma_a\cdot\bsigma_b
  \nonumber\\[-2pt]
  &+A_1\bsigma_a\cdot\bL_2
  +A_2\bL_1\cdot\bsigma_b
  -\bL_1\cdot\bL_2.
  \label{eq:full-axial-kernel}
\end{align}
For example,
\begin{align}
  \bW_1\cdot\bW_2={}&
  \frac{(V_1+M_1)(V_2+M_2)}{4m_N^2}
  \nonumber\\[-2pt]
  &\times\bigl[(\bsigma_a\cdot\bsigma_b)(\bq_1\cdot\bq_2)
  -(\bsigma_a\cdot\bq_2)(\bsigma_b\cdot\bq_1)\bigr].
  \label{eq:magnetism-product}
\end{align}
The axial-charge product contains $AA$, $AT$, $TA$, and $TT$ terms. The remaining axial pieces are the Gamow--Teller, axial--pseudoscalar, and pseudoscalar--pseudoscalar channels. No common-momentum assumption has been made. Setting $S_i=T_i=0$ before forming these products gives the PI current basis in the main text.

The complete ordered-pair operator is obtained from
Eq.~\eqref{eq:PI-two-body-operator} by replacing
$\mathcal K_{X,ab}^{\rm PI}$ with
$\mathcal K_{X,ab}^{\rm full}$.  Under the simultaneous operation
\begin{equation}
  a\leftrightarrow b,
  \qquad \bq\to-\bq,
  \qquad 1\leftrightarrow2,
  \label{eq:exchange-map}
\end{equation}
the two current momenta and all current slots are exchanged.  The ordered $a\ne b$ form is therefore useful for retaining odd harmonics of mixed form-factor products until their accompanying spin and derivative tensors
have also been exchanged.

\section{General multipoles and explicit recoupling}
\label{app:multipoles}

\subsection{Irreducible spin and momentum tensors}

Tensor products and scalar products are defined as
\begin{align}
  [A^{(j)}\otimes B^{(k)}]_{JM}
  &=\sum_{mn}\braket{jm\,kn|JM}A^{(j)}_mB^{(k)}_n,
  \nonumber
  \\
  A^{(j)}\cdot B^{(j)}
  &=\sum_m(-1)^m A^{(j)}_mB^{(j)}_{-m}
  \nonumber
  \\[-2pt]
  &=(-1)^j\hat j [A^{(j)}\otimes B^{(j)}]_{00}.
  \label{eq:tensor-conventions}
\end{align}
Here, the rank is carried by the parenthesized superscript, and the magnetic projection by the subscript. The Cartesian identity
\begin{align}
  \nonumber
  &(\bsigma_a\cdot\mathbf x)(\bsigma_b\cdot\mathbf y)
  \\
  \nonumber
  &=\frac13(\bsigma_a\cdot\bsigma_b)(\mathbf x\cdot\mathbf y)
  +\frac12(\bsigma_a\times\bsigma_b)\cdot(\mathbf x\times\mathbf y)
  \\
  &\qquad+\Sigma^{(2)}_{ab,ij}Q^{(2)}_{ij}(\mathbf x,\mathbf y),
  \label{eq:dyadic-spin-identity}
\end{align}
displays the spin ranks $S=0,1,2$, where
\begin{subequations}
\begin{align}
  &\Sigma^{(2)}_{ab,ij}
  =\frac12(\sigma_{a,i}\sigma_{b,j}+\sigma_{a,j}\sigma_{b,i})
  -\frac13\delta_{ij}\bsigma_a\cdot\bsigma_b,
  \\
  &Q^{(2)}_{ij}(\mathbf x,\mathbf y)
  =\frac12(x_iy_j+x_jy_i)-\frac13\delta_{ij}\mathbf x\cdot\mathbf y.
\end{align}
\end{subequations}
For $\mathbf x=\mathbf y=\bq_i$, the antisymmetric rank-one term vanishes and
\begin{subequations}
\begin{align}
 & (\bsigma_a\cdot\bq_i)(\bsigma_b\cdot\bq_i)
  =\frac{q_i^2}{3}\left[\bsigma_a\cdot\bsigma_b
  +S_{ab}(\widehat\bq_i)\right],
  \\
 & S_{ab}(\widehat{\mathbf n})=3(\bsigma_a\cdot\widehat{\mathbf n})
(\bsigma_b\cdot\widehat{\mathbf n})-\bsigma_a\cdot\bsigma_b.
\end{align}
  \label{eq:AP-spin-tensor}
\end{subequations}
Equation \eqref{eq:dyadic-spin-identity} also decomposes the $PP$, $WW$ and axial-charge channels.

With $s_1=+1$ and $s_2=-1$, the routing in
Eq.~\eqref{eq:momentum-routing} reads
\begin{equation}
  \bq_i=-\frac{p}{2} C_1(\widehat\bp)
  +s_iq C_1(\widehat\bq).
  \label{eq:q-slot-spherical}
\end{equation}
It implies
\begin{equation}
  \bq_1\cdot\bq_2=\frac{p^2}{4}-q^2,
  \qquad
  \bq_1\times\bq_2=\bp\times\bq.
  \label{eq:q-dot-cross}
\end{equation}
Every explicit momentum tensor in Eq.~\eqref{eq:PI-channels} is generated by coupling the two terms in Eq.~\eqref{eq:q-slot-spherical}; its rank is at most two. Derivative channels are obtained by replacing one or both current momenta with $\bK_a$ or $\bK_b$.

\subsection{Complete PI primitive dictionary}
\label{app:primitive-dictionary}

Now we make the primitive label explicit. We define the orbital tensors
\begin{align}
\nonumber
U_{ij,S}&\equiv[\bq_i^{(1)}\otimes\bq_j^{(1)}]_S
\\
\nonumber
&=\sum_{l_pl_q}u^{ij,S}_{l_pl_q}(p,q)
[C_{l_p}(\widehat\bp)\otimes C_{l_q}(\widehat\bq)]_S,
\\
\bq_i^{(1)}&=\sum_{l_pl_q}v^i_{l_pl_q}(p,q)
[C_{l_p}(\widehat\bp)\otimes C_{l_q}(\widehat\bq)]_1 ,
\label{eq:uv-definition}
\end{align}
where the only nonzero coefficients for $u$ and $v$ are
\begin{align}
\nonumber
&u^{ij,0}_{00}=-\frac{1}{\sqrt3}\left(\frac{p^2}{4}+s_is_jq^2\right),\quad
u^{ij,0}_{11}=-\frac{pq}{2}(s_i+s_j),
\\
\nonumber
&u^{ij,1}_{11}=-\frac{pq}{2}(s_j-s_i),
\\
\nonumber
&u^{ij,2}_{20}=\sqrt{\frac23}\frac{p^2}{4},\quad
u^{ij,2}_{02}=\sqrt{\frac23}s_is_jq^2,
\\
&u^{ij,2}_{11}=-\frac{pq}{2}(s_i+s_j),
\quad v^i_{10}=-\frac p2.
\quad v^i_{01}=s_iq.
\label{eq:uv-coefficients}
\end{align}
The above is a complete list of every allowed pair $(l_p,l_q)$. It implies $U_{ii,1}=0$ and $U_{12,1}=pq[C_1(\widehat\bp)\otimes C_1(\widehat\bq)]_1$.

The independent spin-derivative tensors $\Xi_{\rho S}$ include
\begin{align}
\nonumber
&\Sigma_{S}=[\bsigma_a^{(1)}\otimes\bsigma_b^{(1)}]_S,\quad S=0,1,2,
\\
\nonumber
&D_{ab,0}=[\bK_a^{(1)}\otimes\bK_b^{(1)}]_0,
\\
\nonumber
&Z_{ab,0}=(\bsigma_a\cdot\bK_a) (\bsigma_b\cdot\bK_b),
\\
\nonumber
&D_{b|a,1}=[\bK_b^{(1)}\otimes\bsigma_a^{(1)}]_1,
\\
&D_{a|b,1}=[\bK_a^{(1)}\otimes\bsigma_b^{(1)}]_1,
\label{eq:Xi-dictionary}
\end{align}
as well as ${\bf 1}$.
Employing Eqs.~\eqref{eq:uv-definition}, \eqref{eq:uv-coefficients}, and \eqref{eq:Xi-dictionary} in \eqref{eq:PI-channels} yields \cref{tab:PI-primitive-dictionary} in which 
each row corresponds to one primitive in Eq.~\eqref{eq:primitive-channel}, with definite $S,l_p,l_q$ numbers, a full weight $w_\rho(p,q)$, and a nuclear tensor $\Xi_{\rho S}$.
\begin{table*}[t]
\centering
\small
\setlength{\tabcolsep}{6pt}
\setlength{\arrayrulewidth}{0.3pt}
\renewcommand{\arraystretch}{1.3}
\begin{tabular}{@{}c|cccccc@{}}
\noalign{\hrule height 0.8pt}
Channel & $XY$ & $S$ & $l_p$ & $l_q$ & $w_\rho(p,q)$ & $\Xi_{\rho S}$
\\
\noalign{\hrule height 0.5pt}
$VV$ & $VV$ & $0$ & $0$ & $0$ & $C_{\Gamma V}$ & $1$
\\
\hline
$BB$ & $VV$ & $0$ & $0$ & $0$ & $\sqrt3C_{\Gamma V}/(4m_N^2)$ & $D_{ab,0}$\\
\hline
\multirow{4}{*}{$WW$} & \multirow{4}{*}{$GG$} & $0$ & $0$ & $0$
& $-C_{\Gamma V}(p^2/4-q^2)/(2\sqrt3m_N^2)$ & $\Sigma_0$\\
& & $1$ & $1$ & $1$
& $\sqrt3C_{\Gamma V} pq/(4m_N^2)$ & $\Sigma_1$\\
& & $2$ & $2$ & $0$
& $-\sqrt{10/3}C_{\Gamma V} p^2/(16m_N^2)$ & $\Sigma_2$\\
& & $2$ & $0$ & $2$
& $\sqrt{10/3}C_{\Gamma V} q^2/(4m_N^2)$ & $\Sigma_2$\\
\hline
\multirow{4}{*}{$BW$} & \multirow{2}{*}{$VG$} & $1$ & $1$ & $0$
& $-\sqrt6C_{\Gamma V} p/(8m_N^2)$ & $D_{a|b,1}$\\
& & $1$ & $0$ & $1$
& $-\sqrt6C_{\Gamma V} q/(4m_N^2)$ & $D_{a|b,1}$\\
\cline{2-7}
& \multirow{2}{*}{$GV$} & $1$ & $1$ & $0$
& $-\sqrt6C_{\Gamma V} p/(8m_N^2)$ & $D_{b|a,1}$\\
& & $1$ & $0$ & $1$
& $\sqrt6C_{\Gamma V} q/(4m_N^2)$ & $D_{b|a,1}$
\\
\hline
$AA$ & $AA$ & $0$ & $0$ & $0$ & $\sqrt3C_{\Gamma A}$ & $\Sigma_0$
\\
\hline
$A^0A^0$ & $AA$ & $0$ & $0$ & $0$ & $C_{\Gamma A}/(4m_N^2)$ & $Z_{ab,0}$
\\
\hline
\multirow{10}{*}{$AP$} & \multirow{5}{*}{$AP$} & $0$ & $0$ & $0$
& $-C_{\Gamma A}(p^2/4+q^2)/(4\sqrt3m_N^2)$ & $\Sigma_0$\\
& & $0$ & $1$ & $1$
& $C_{\Gamma A} pq/(4m_N^2)$ & $\Sigma_0$\\
& & $2$ & $2$ & $0$
& $\sqrt{10/3}C_{\Gamma A} p^2/(16m_N^2)$ & $\Sigma_2$\\
& & $2$ & $0$ & $2$
& $\sqrt{10/3}C_{\Gamma A} q^2/(4m_N^2)$ & $\Sigma_2$\\
& & $2$ & $1$ & $1$
& $\sqrt5C_{\Gamma A} pq/(4m_N^2)$ & $\Sigma_2$\\
\cline{2-7}
& \multirow{5}{*}{$PA$} & $0$ & $0$ & $0$
& $-C_{\Gamma A}(p^2/4+q^2)/(4\sqrt3m_N^2)$ & $\Sigma_0$\\
& & $0$ & $1$ & $1$
& $-C_{\Gamma A} pq/(4m_N^2)$ & $\Sigma_0$\\
& & $2$ & $2$ & $0$
& $\sqrt{10/3}C_{\Gamma A} p^2/(16m_N^2)$ & $\Sigma_2$\\
& & $2$ & $0$ & $2$
& $\sqrt{10/3}C_{\Gamma A} q^2/(4m_N^2)$ & $\Sigma_2$\\
& & $2$ & $1$ & $1$
& $-\sqrt5C_{\Gamma A} pq/(4m_N^2)$ & $\Sigma_2$\\
\hline
\multirow{4}{*}{$PP$} & \multirow{4}{*}{$PP$} & $0$ & $0$ & $0$
& $C_{\Gamma A}(p^2/4-q^2)^2/(16\sqrt3m_N^4)$ & $\Sigma_0$\\
& & $1$ & $1$ & $1$
& $-\sqrt3C_{\Gamma A}(p^2/4-q^2)pq/(16m_N^4)$ & $\Sigma_1$\\
& & $2$ & $2$ & $0$
& $-\sqrt{10/3}C_{\Gamma A}(p^2/4-q^2)p^2/(64m_N^4)$ & $\Sigma_2$\\
& & $2$ & $0$ & $2$
& $\sqrt{10/3}C_{\Gamma A}(p^2/4-q^2)q^2/(16m_N^4)$ & $\Sigma_2$\\
\noalign{\hrule height 0.8pt}
\end{tabular}
\caption{Fully expanded primitive dictionary for the eight PI channels: one row for each of the 26 nonzero elementary tensor terms at generic $p,q$. Each row specifies $S,l_p,l_q$, the full weight $w_\rho(p,q)$, and the nuclear tensor $\Xi_{\rho S}$ in \cref{eq:primitive-channel}; its contribution to Eq.~\eqref{eq:PI-channels} is $X(Q_1^2)Y(Q_2^2)w_\rho(p,q) [[C_{l_p}(\widehat\bp)\otimes C_{l_q}(\widehat\bq)]_S\otimes\Xi_{\rho S}]_{00}$.
The labels $l_p,l_q$ refer to the explicit momentum harmonics before expanding the form-factor angular dependence. Unlisted $(S,l_p,l_q)$ combinations have zero weight.}
\label{tab:PI-primitive-dictionary}
\end{table*}

\subsection{Form-factor recoupling and radial kernels}

Addition theorem gives
\begin{align}
  P_n(\widehat\bp\cdot\widehat\bq)
  =(-1)^n\hat n
  [C_n(\widehat\bp)\otimes C_n(\widehat\bq)]_{00}.
  \label{eq:Legendre-C}
\end{align}
For a primitive tensor of the form in Eq.~\eqref{eq:primitive-channel}, the $n$th form-factor harmonic recouples according to
\begin{align}
  &P_n(z)\left[[C_{l_p}(\widehat\bp)\otimes
  C_{l_q}(\widehat\bq)]_S\otimes\Xi_{\rho S}\right]_{00}
  \nonumber\\[-2pt]
  =&\sum_{L_pL_q}\mathcal R^S_n(l_pl_q;L_pL_q)
  \left[[C_{L_p}(\widehat\bp)\otimes
  C_{L_q}(\widehat\bq)]_S\otimes\Xi_{\rho S}\right]_{00},
  \label{eq:FF-recoupling}
\end{align}
where
\begin{align}
  \mathcal R^S_n(l_pl_q;L_pL_q)
  ={}&(-1)^n\hat n \hat S \hat L_p \hat L_q
  \begin{Bmatrix}
    n&l_p&L_p\\ n&l_q&L_q\\ 0&S&S
  \end{Bmatrix}
  \nonumber\\[-2pt]
  &\times
  \braket{n0\,l_p0|L_p0}
  \braket{n0\,l_q0|L_q0}.
  \label{eq:nine-j-coefficient}
\end{align}
The factor $(-1)^n\hat n$ follows from $P_n=(-1)^n\hat n[C_n(\widehat\bp)\otimes C_n(\widehat\bq)]_{00}$.  As a normalization check,
$\mathcal R^S_0(l_pl_q;L_pL_q)=\delta_{L_p l_p}\delta_{L_q l_q}$. The $9j$ symbol and Clebsch--Gordan coefficients enforce all triangle and parity conditions.

The base ranks in \cref{tab:PI-primitive-dictionary} are followed by the form-factor harmonic $n$. The allowed recoupled ranks are
  \begin{align}
    \nonumber
    &|n-l_p|\le L_p\le n+l_p,\quad n+l_p+L_p\ \hbox{even},
    \\
    \nonumber
    &|n-l_q|\le L_q\le n+l_q,\quad n+l_q+L_q\ \hbox{even},
    \\
    &|L_p-L_q|\le S\le L_p+L_q.
    \label{eq:allowed-angular-ranks}
  \end{align}
Equal products $VV,GG,AA,PP$ contain only even $n$.  For the mixed products, \cref{eq:mixed-FF-exchange} and the paired $VG/GV$ or $AP/PA$ families enforce exchange symmetry.  In the scalar operator one further has $L_p=L$ from \cref{eq:zero-CG}, where $L$ is introduced in \cref{eq:center-C-expansion}. Thus \cref{eq:allowed-angular-ranks}, together with the table, is the complete set of angular restrictions.

For a channel with an ordered form-factor product $XY$, we introduce
\begin{align}
  \nonumber
  &\mathcal H^\rho_{L_pL_qS}(r;p)
  \\[-2pt]
  \nonumber
  =&4\pi i^{L_q}\int_0^\infty\frac{q^2\dd q}{(2\pi)^3} j_{L_q}(qr)
  \sum_n \hat n^2\mathcal F_n^{XY}(p,q)
  \\[-2pt]
  &\times w_\rho(p,q)
  \mathcal R^S_n(l_pl_q;L_pL_q).
  \label{eq:radial-H}
\end{align}
The phase follows from the convention in \cref{eq:two-current-phase},
\begin{align}
  &\int\dd\Omega_q\,e^{i\bq\cdot\br}C_{L_qm}(\widehat\bq)
  =4\pi i^{L_q}j_{L_q}(qr)C_{L_qm}(\widehat\br).
  \label{eq:Fourier-C}
\end{align}
No angular integration remains in \cref{eq:radial-H}.

\subsection{Pair-center phase and arbitrary nuclear rank}

For the phase associated with the center-of-mass coordinate of a nucleon pair, we expand as usual:
\begin{align}
  e^{-i\bp\cdot\bR}=\sum_L\hat L^2
  (-i)^Lj_L(pR)
  C_L(\widehat\bp)\cdot C_L(\widehat\bR).
  \label{eq:center-C-expansion}
\end{align}
An explicit magnetic-sum form of the tensor accompanying $C^*_{JM}(\widehat\bp)$ in the convention of \cref{eq:PI-multipole-expansion} is
\begin{widetext}
  \begin{align}
    \mathfrak T^{\rho;JM}_{L,L_p,L_q,S}
    \equiv{}&\braket{L0\,L_p0|J0}
    \sum_{m m_p m_q\mu M_S}(-1)^{m+M}
    \braket{L_pm_p\,L_qm_q|SM_S}
    \nonumber\\[-2pt]
    &\times\braket{SM_S\,S\mu|00}
    \braket{Lm\,L_pm_p|J,-M}
    C_{L,-m}(\widehat\bR)C_{L_qm_q}(\widehat\br)
    \Xi_{\rho S\mu}.
    \label{eq:final-magnetic-tensor}
  \end{align}
  The phase $(-1)^M$ and the projection $-M$ are required when the product of the two $\widehat\bp$ harmonics is converted from its $C_{J,-M}$ component to the coefficient of $C^*_{JM}$. For $J=M=0$ they reduce to unity, so the scalar collapse below is unchanged. The general multipole is consequently
  \begin{align}
    \widetilde\cO_{\Gamma;JM}^{\rm full}(p)
    ={}&\sum_{a\ne b}\tau_a^-\tau_b^-
    \sum_{\rho,L,L_p,L_q,S}\hat L^2
    (-i)^Lj_L(pR_{ab}) \mathcal H^\rho_{L_pL_qS}(r_{ab};p) \mathfrak T^{\rho;JM}_{L,L_p,L_q,S}(a,b).
    \label{eq:general-complete-multipole}
  \end{align}
\end{widetext}
For the six-form-factor current, $\rho$ runs over every primitive term generated by \cref{eq:full-vector-kernel} and \cref{eq:full-axial-kernel}. For the PI current, it is restricted to \cref{eq:PI-channel-set}. Setting $J=0$ in \cref{eq:final-magnetic-tensor} gives
\begin{align}
  \mathfrak T^{\rho;00}_{L,L_p,L_q,S}
  =\frac{\delta_{L_pL}}{\hat L^2}
  \left[[C_L(\widehat\bR)\otimes C_{L_q}(\widehat\br)]_S
  \otimes\Xi_{\rho S}\right]_{00},
  \label{eq:scalar-tensor-collapse}
\end{align}
which proves Eq.~\eqref{eq:PI-zero-plus-operator}.

\section{Power-counting qualifications}
\label{app:checks}

The Darwin and spin--orbit corrections mentioned at the end of \cref{sec:PI-currents} do not require an explicit coordinate in the covariant nucleon vertex.  They arise when the momentum-space Dirac bilinear is expanded between Pauli spinors and then represented on nuclear wave functions. For example, through the next nonvanishing order, the vector charge contains schematically
\begin{align}
  \nonumber
  \mathcal V_i^0&=V_i-\frac{V_i+2M_i}{8m_N^2}Q_i^2
  \\
  &\qquad+\frac{i(V_i+2M_i)}{8m_N^2}\bsigma_i\cdot(\bq_i\times\bK_i)+\cdots,
  \label{eq:Darwin-spin-orbit}
\end{align}
apart from convention-dependent reshuffling associated with external-state normalization. The $Q_i^2$ term is the Darwin--Foldy charge correction; the next term is the momentum-space spin--orbit charge. After Fourier transformation,
\begin{align}
  Q^2\widetilde f(\bq)&\longleftrightarrow-\nabla^2f(\br),
  \nonumber\\[-2pt]
  i\bsigma\cdot(\bq\times\bK)\widetilde f(\bq)
  &\longleftrightarrow
  \bsigma\cdot[\boldsymbol\nabla f\times
  (-i\overleftrightarrow{\boldsymbol\nabla})].
  \label{eq:Darwin-coordinate-map}
\end{align}
Here $f(\br)=\int\dd^3q\,e^{i\bq\cdot\br}\widetilde f(\bq)/(2\pi)^3$ is the coordinate-space transform of the momentum-space kernel. The ordinary gradient $\boldsymbol\nabla f$ differentiates only this kernel. By contrast, $\overleftrightarrow{\boldsymbol\nabla}$ is the two-sided nucleon derivative defined in \cref{eq:nucleon-average-momentum}: its right-acting part acts on the nuclear ket and its left-acting part acts on the nuclear bra. For a central kernel, the second structure contains
$[f'(r)/r]\bsigma\cdot\mathbf L$.  The coordinate is supplied by the Fourier transform and the nuclear wave function, not inserted at the local vertex.

These corrections interfere with the leading charge and therefore occur at the same explicit order as $BB$, $WW$, and $BW$.  Analogous non-pole axial recoil corrections interfere with $A_i\bsigma_i$. This is why the PI kernel in the main text is described as a selected-current reduction rather than a complete $O(m_N^{-2})$ calculation.

\begin{acknowledgments}
  This work was supported in part by Grants No.\,NSFC-12605178 and
  No.\,NSFC-12035008.
  We are grateful to Profs. Dong-Liang Fang, Jiang-Ming Yao, and Peng-Wei Zhao for reading our manuscript and making helpful comments.
  YL thanks JL for support.
\end{acknowledgments}

\bibliographystyle{utphys}
\bibliography{references}

@article{Sakharov1967,
  author = {Sakharov, A. D.},
  title = {Violation of CP Invariance, C Asymmetry, and Baryon Asymmetry of the Universe},
  journal = {JETP Lett.},
  volume = {5},
  pages = {24--27},
  year = {1967}
}

@article{FukugitaYanagida1986,
  author = {Fukugita, M. and Yanagida, T.},
  title = {Baryogenesis Without Grand Unification},
  journal = {Phys. Lett. B},
  volume = {174},
  pages = {45--47},
  year = {1986},
  doi = {10.1016/0370-2693(86)91126-3}
}

@article{Minkowski1977,
  author = {Minkowski, Peter},
  title = {Muon to Electron Gamma at a Rate of One Out of One Billion Muon Decays?},
  journal = {Phys. Lett. B},
  volume = {67},
  pages = {421--428},
  year = {1977},
  doi = {10.1016/0370-2693(77)90435-X}
}

@inproceedings{Yanagida1979,
  author = {Yanagida, Tsutomu},
  title = {Horizontal Symmetry and Masses of Neutrinos},
  booktitle = {Proceedings of the Workshop on the Unified Theory and the Baryon Number in the Universe},
  editor = {Sawada, O. and Sugamoto, A.},
  pages = {95--99},
  year = {1979},
  publisher = {KEK}
}

@inproceedings{GellMannRamondSlansky1979,
  author = {Gell-Mann, Murray and Ramond, Pierre and Slansky, Richard},
  title = {{Complex Spinors and Unified Theories}},
  booktitle = {Supergravity},
  editor = {van Nieuwenhuizen, Peter and Freedman, Daniel Z.},
  pages = {315--321},
  year = {1979},
  publisher = {North-Holland},
  eprint = {1306.4669},
  archivePrefix = {arXiv},
  primaryClass = {hep-th}
}

@incollection{Glashow1980,
  author = {Glashow, Sheldon L.},
  title = {{The Future of Elementary Particle Physics}},
  booktitle = {Quarks and Leptons},
  editor = {L{\'e}vy, Maurice and Basdevant, Jean-Louis and Speiser, David and Weyers, Jacques and Gastmans, Raymond and Jacob, Maurice},
  pages = {687--713},
  year = {1980},
  publisher = {Plenum Press},
  address = {New York},
  doi = {10.1007/978-1-4684-7197-7_15}
}

@article{MohapatraSenjanovic1980,
  author = {Mohapatra, Rabindra N. and Senjanovi{\'c}, Goran},
  title = {{Neutrino Mass and Spontaneous Parity Nonconservation}},
  journal = {Phys. Rev. Lett.},
  volume = {44},
  pages = {912--915},
  year = {1980},
  doi = {10.1103/PhysRevLett.44.912}
}

@article{Weinberg1979,
  author = {Weinberg, Steven},
  title = {Baryon- and Lepton-Nonconserving Processes},
  journal = {Phys. Rev. Lett.},
  volume = {43},
  pages = {1566--1570},
  year = {1979},
  doi = {10.1103/PhysRevLett.43.1566}
}

@article{SchechterValle1982,
  author = {Schechter, J. and Valle, J. W. F.},
  title = {Neutrinoless Double-Beta Decay in SU(2) x U(1) Theories},
  journal = {Phys. Rev. D},
  volume = {25},
  pages = {2951--2954},
  year = {1982},
  doi = {10.1103/PhysRevD.25.2951}
}

@article{EngelMenendez2017,
  author = {Engel, Jonathan and Menendez, Javier},
  title = {Status and Future of Nuclear Matrix Elements for Neutrinoless Double-Beta Decay: A Review},
  journal = {Rep. Prog. Phys.},
  volume = {80},
  number = {4},
  pages = {046301},
  year = {2017},
  doi = {10.1088/1361-6633/aa5bc5}
}

@article{Dolinski2019,
  author = {Dolinski, Michelle J. and Poon, Alan W. P. and Rodejohann, Werner},
  title = {Neutrinoless Double-Beta Decay: Status and Prospects},
  journal = {Annu. Rev. Nucl. Part. Sci.},
  volume = {69},
  pages = {219--251},
  year = {2019},
  doi = {10.1146/annurev-nucl-101918-023407}
}

@article{Dohmen1993,
  author = {Dohmen, C. and others},
  title = {Test of Lepton-Flavor Conservation in Muon-to-Electron Conversion on Titanium},
  journal = {Phys. Lett. B},
  volume = {317},
  pages = {631--636},
  year = {1993},
  doi = {10.1016/0370-2693(93)91383-X}
}

@article{Kaulard1998,
  author = {Kaulard, J. and others},
  title = {Improved Limit on the Branching Ratio of Muon-to-Positron Conversion on Titanium},
  journal = {Phys. Lett. B},
  volume = {422},
  pages = {334--338},
  year = {1998},
  doi = {10.1016/S0370-2693(97)01423-8}
}

@techreport{Mu2eTDR,
  author = {Bartoszek, L. and others},
  collaboration = {Mu2e},
  title = {Mu2e Technical Design Report},
  institution = {Fermi National Accelerator Laboratory},
  number = {FERMILAB-TM-2594},
  year = {2014},
  doi = {10.2172/1172555},
  eprint = {1501.05241},
  archivePrefix = {arXiv},
  primaryClass = {physics.ins-det}
}

@article{COMETTDR,
  author = {Abramishvili, R. and others},
  collaboration = {COMET},
  title = {COMET Phase-I Technical Design Report},
  journal = {Prog. Theor. Exp. Phys.},
  volume = {2020},
  number = {3},
  pages = {033C01},
  year = {2020},
  doi = {10.1093/ptep/ptz125}
}

@article{Traini2001,
  author = {Traini, M.},
  title = {Coulomb Distortion in Quasielastic $(e,e')$ Scattering on Nuclei: Effective Momentum Approximation and Beyond},
  journal = {Nucl. Phys. A},
  volume = {694},
  pages = {325--336},
  year = {2001},
  doi = {10.1016/S0375-9474(01)00968-X}
}

@article{WallaceTjon2008,
  author = {Wallace, S. J. and Tjon, J. A.},
  title = {Coulomb Corrections in Quasi-Elastic Scattering: Tests of the Effective-Momentum Approximation},
  journal = {Phys. Rev. C},
  volume = {78},
  pages = {044604},
  year = {2008},
  doi = {10.1103/PhysRevC.78.044604}
}

@article{Haxton2023,
  author = {Haxton, W. C. and Rule, E. and McElvain, K. and Ramsey-Musolf, M. J.},
  title = {Nuclear-Level Effective Theory of Muon-to-Electron Conversion: Formalism and Applications},
  journal = {Phys. Rev. C},
  volume = {107},
  pages = {035504},
  year = {2023},
  doi = {10.1103/PhysRevC.107.035504}
}

@article{HaxtonRule2025,
  author = {Haxton, W. C. and Rule, E.},
  title = {Nuclear-Level Effective Theory of Muon-to-Electron Conversion: Inelastic Process},
  journal = {Phys. Rev. C},
  volume = {111},
  pages = {025501},
  year = {2025},
  doi = {10.1103/PhysRevC.111.025501}
}

@article{BabuLeung2001,
  author = {Babu, K. S. and Leung, C. N.},
  title = {Classification of Effective Neutrino Mass Operators},
  journal = {Nucl. Phys. B},
  volume = {619},
  pages = {667--689},
  year = {2001},
  doi = {10.1016/S0550-3213(01)00504-1},
  eprint = {hep-ph/0106054},
  archivePrefix = {arXiv}
}

@article{deGouveaJenkins2008,
  author = {de Gouv{\^e}a, Andr{\'e} and Jenkins, James},
  title = {A Survey of Lepton Number Violation via Effective Operators},
  journal = {Phys. Rev. D},
  volume = {77},
  pages = {013008},
  year = {2008},
  doi = {10.1103/PhysRevD.77.013008},
  eprint = {0708.1344},
  archivePrefix = {arXiv},
  primaryClass = {hep-ph}
}

@article{Lehman2014,
  author = {Lehman, Landon},
  title = {Extending the Standard Model Effective Field Theory with the Complete Set of Dimension-7 Operators},
  journal = {Phys. Rev. D},
  volume = {90},
  pages = {125023},
  year = {2014},
  doi = {10.1103/PhysRevD.90.125023},
  eprint = {1410.4193},
  archivePrefix = {arXiv},
  primaryClass = {hep-ph}
}

@article{LiaoMa2016Dim7,
  author = {Liao, Yi and Ma, Xiao-Dong},
  title = {Renormalization Group Evolution of Dimension-Seven Baryon- and Lepton-Number-Violating Operators},
  journal = {J. High Energy Phys.},
  volume = {2016},
  number = {11},
  pages = {043},
  year = {2016},
  doi = {10.1007/JHEP11(2016)043},
  eprint = {1607.07309},
  archivePrefix = {arXiv},
  primaryClass = {hep-ph}
}

@article{LiaoMa2019Dim7,
  author = {Liao, Yi and Ma, Xiao-Dong},
  title = {Renormalization Group Evolution of Dimension-Seven Operators in Standard Model Effective Field Theory and Relevant Phenomenology},
  journal = {J. High Energy Phys.},
  volume = {2019},
  number = {03},
  pages = {179},
  year = {2019},
  doi = {10.1007/JHEP03(2019)179},
  eprint = {1901.10302},
  archivePrefix = {arXiv},
  primaryClass = {hep-ph}
}

@article{LiaoMa2020,
  author = {Liao, Yi and Ma, Xiao-Dong},
  title = {An Explicit Construction of the Dimension-9 Operator Basis in the Standard Model Effective Field Theory},
  journal = {J. High Energy Phys.},
  volume = {2020},
  number = {11},
  pages = {152},
  year = {2020},
  doi = {10.1007/JHEP11(2020)152},
  eprint = {2007.08125},
  archivePrefix = {arXiv},
  primaryClass = {hep-ph}
}

@article{Li:2020xlh,
    author = "Li, Hao-Lin and Ren, Zhe and Xiao, Ming-Lei and Yu, Jiang-Hao and Zheng, Yu-Hui",
    title = "{Complete set of dimension-nine operators in the standard model effective field theory}",
    eprint = "2007.07899",
    archivePrefix = "arXiv",
    primaryClass = "hep-ph",
    doi = "10.1103/PhysRevD.104.015025",
    journal = "Phys. Rev. D",
    volume = "104",
    number = "1",
    pages = "015025",
    year = "2021"
}

@article{LiaoMaWang2020,
  author = {Liao, Yi and Ma, Xiao-Dong and Wang, Quan-Yu},
  title = {Extending Low Energy Effective Field Theory with a Complete Set of Dimension-7 Operators},
  journal = {J. High Energy Phys.},
  volume = {2020},
  number = {08},
  pages = {162},
  year = {2020},
  doi = {10.1007/JHEP08(2020)162},
  eprint = {2005.08013},
  archivePrefix = {arXiv},
  primaryClass = {hep-ph}
}

@article{LiEtAl2021,
  author = {Li, Hao-Lin and Ren, Zhe and Xiao, Ming-Lei and Yu, Jiang-Hao and Zheng, Yu-Hui},
  title = {Low Energy Effective Field Theory Operator Basis at $d\leq 9$},
  journal = {J. High Energy Phys.},
  volume = {2021},
  number = {06},
  pages = {138},
  year = {2021},
  doi = {10.1007/JHEP06(2021)138},
  eprint = {2012.09188},
  archivePrefix = {arXiv},
  primaryClass = {hep-ph}
}

@article{Graesser2017,
  author = {Graesser, Michael L.},
  title = {An Electroweak Basis for Neutrinoless Double-Beta Decay},
  journal = {J. High Energy Phys.},
  volume = {2017},
  number = {08},
  pages = {099},
  year = {2017},
  doi = {10.1007/JHEP08(2017)099},
  eprint = {1606.04549},
  archivePrefix = {arXiv},
  primaryClass = {hep-ph}
}

@article{LiaoEtAl2020,
  author = {Liao, Yi and Ma, Xiao-Dong and Wang, Hao-Lin},
  title = {Effective Field Theory Approach to Lepton Number Violating Decays $K^\pm\to\pi^\mp\ell^\pm\ell^\pm$: Short-Distance Contribution},
  journal = {J. High Energy Phys.},
  volume = {2020},
  number = {01},
  pages = {127},
  year = {2020},
  doi = {10.1007/JHEP01(2020)127},
  eprint = {1909.06272},
  archivePrefix = {arXiv},
  primaryClass = {hep-ph}
}

@article{Cirigliano2018Master,
  author = {Cirigliano, Vincenzo and Dekens, Wouter and de Vries, Jordy and Graesser, Michael L. and Mereghetti, Emanuele},
  title = {A Neutrinoless Double-Beta Decay Master Formula from Effective Field Theory},
  journal = {J. High Energy Phys.},
  volume = {2018},
  number = {12},
  pages = {097},
  year = {2018},
  doi = {10.1007/JHEP12(2018)097},
  eprint = {1806.02780},
  archivePrefix = {arXiv},
  primaryClass = {hep-ph}
}

@article{Cirigliano2018Contact,
  author = {Cirigliano, Vincenzo and Dekens, Wouter and de Vries, Jordy and Graesser, Michael L. and Mereghetti, Emanuele and Pastore, Saori and van Kolck, Ubirajara},
  title = {{A New Leading Contribution to Neutrinoless Double-Beta Decay}},
  journal = {Phys. Rev. Lett.},
  volume = {120},
  pages = {202001},
  year = {2018},
  doi = {10.1103/PhysRevLett.120.202001},
  eprint = {1802.10097},
  archivePrefix = {arXiv},
  primaryClass = {nucl-th}
}

@article{Yao2015,
  author = {Yao, Jiang-Ming and Song, L. S. and Hagino, K. and Ring, P. and Meng, J.},
  title = {{Systematic Study of Nuclear Matrix Elements in Neutrinoless Double-Beta Decay with a Beyond-Mean-Field Covariant Density Functional Theory}},
  journal = {Phys. Rev. C},
  volume = {91},
  pages = {024316},
  year = {2015},
  doi = {10.1103/PhysRevC.91.024316},
  eprint = {1410.6326},
  archivePrefix = {arXiv},
  primaryClass = {nucl-th}
}

@article{WangZhaoMeng2021,
  author = {Wang, Y. K. and Zhao, P. W. and Meng, J.},
  title = {{Nuclear Matrix Elements of Neutrinoless Double-Beta Decay in the Triaxial Projected Shell Model}},
  journal = {Phys. Rev. C},
  volume = {104},
  pages = {014320},
  year = {2021},
  doi = {10.1103/PhysRevC.104.014320},
  eprint = {2105.02649},
  archivePrefix = {arXiv},
  primaryClass = {nucl-th}
}

@article{FangFaesslerSimkovic2018,
  author = {Fang, Dong-Liang and Faessler, Amand and {\v S}imkovic, Fedor},
  title = {{$0\nu\beta\beta$-Decay Nuclear Matrix Elements for Light and Heavy Neutrino-Mass Mechanisms from Deformed QRPA Calculations with Isospin Restoration}},
  journal = {Phys. Rev. C},
  volume = {97},
  pages = {045503},
  year = {2018},
  doi = {10.1103/PhysRevC.97.045503},
  eprint = {1803.09195},
  archivePrefix = {arXiv},
  primaryClass = {nucl-th}
}

@article{Simkovic2001,
  author = {{\v S}imkovic, Fedor and Domin, Pavol and Kovalenko, Sergey and Faessler, Amand},
  title = {The Muon-Positron Conversion in Nuclei Mediated by Light Majorana Neutrinos},
  journal = {Part. Nucl. Lett.},
  volume = {104},
  pages = {40--55},
  year = {2001},
  eprint = {hep-ph/0103029},
  archivePrefix = {arXiv}
}

@article{Divari2002,
  author = {Divari, P. C. and Vergados, J. D. and Kosmas, T. S. and Skouras, L. D.},
  title = {The Exotic Double-Charge-Exchange Muon-Positron Conversion in Nuclei},
  journal = {Nucl. Phys. A},
  volume = {703},
  pages = {409--431},
  year = {2002},
  doi = {10.1016/S0375-9474(01)01533-0},
  eprint = {nucl-th/0203066},
  archivePrefix = {arXiv}
}

@article{Domin2004,
  author = {Domin, Pavol and Kovalenko, Sergey and Faessler, Amand and {\v S}imkovic, Fedor},
  title = {Nuclear Muon-Positron Conversion Mediated by Majorana Neutrinos},
  journal = {Phys. Rev. C},
  volume = {70},
  pages = {065501},
  year = {2004},
  doi = {10.1103/PhysRevC.70.065501},
  eprint = {nucl-th/0409033},
  archivePrefix = {arXiv}
}

@article{GeibMerleZuber2017,
  author = {Geib, Tanja and Merle, Alexander and Zuber, Kai},
  title = {$\mu^-$--$e^+$ Conversion in Upcoming Lepton-Flavor-Violation Experiments},
  journal = {Phys. Lett. B},
  volume = {764},
  pages = {157--162},
  year = {2017},
  doi = {10.1016/j.physletb.2016.11.029},
  eprint = {1609.09088},
  archivePrefix = {arXiv},
  primaryClass = {hep-ph}
}

@article{Geib:2016daa,
    author = "Geib, Tanja and Merle, Alexander",
    title = "{$\mu^-$--$e^+$ Conversion from Short-Range Operators}",
    eprint = "1612.00452",
    archivePrefix = "arXiv",
    primaryClass = "hep-ph",
    doi = "10.1103/PhysRevD.95.055009",
    journal = "Phys. Rev. D",
    volume = "95",
    number = "5",
    pages = "055009",
    year = "2017"
}

@article{Berryman2017,
  author = {Berryman, Jeffrey M. and de Gouv{\^e}a, Andr{\'e} and Kelly, Kevin J. and Kobach, Andrew},
  title = {Lepton-Number-Violating Searches for Muon-to-Positron Conversion},
  journal = {Phys. Rev. D},
  volume = {95},
  pages = {115010},
  year = {2017},
  doi = {10.1103/PhysRevD.95.115010},
  eprint = {1611.00032},
  archivePrefix = {arXiv},
  primaryClass = {hep-ph}
}

@article{LeeMacKenzie2022,
  author = {Lee, MyeongJae and MacKenzie, Michael},
  title = {Muon-to-Positron Conversion},
  journal = {Universe},
  volume = {8},
  number = {4},
  pages = {227},
  year = {2022},
  doi = {10.3390/universe8040227},
  eprint = {2110.07093},
  archivePrefix = {arXiv},
  primaryClass = {hep-ex}
}

@article{WeinbergFeinberg1959,
  author = {Weinberg, Steven and Feinberg, Gerald},
  title = {Electromagnetic Transitions between Muon Meson and Electron},
  journal = {Phys. Rev. Lett.},
  volume = {3},
  pages = {111--114},
  year = {1959},
  doi = {10.1103/PhysRevLett.3.111}
}

@article{Kosmas1994,
  author = {Kosmas, T. S. and Vergados, J. D.},
  title = {Muon-to-Electron Conversion: A Symbiosis of Particle and Nuclear Physics},
  journal = {Phys. Rep.},
  volume = {264},
  pages = {251--266},
  year = {1996},
  doi = {10.1016/0370-1573(95)00041-0},
  eprint = {nucl-th/9408011},
  archivePrefix = {arXiv}
}

@article{KitanoKoikeOkada2002,
  author = {Kitano, Ryuichiro and Koike, Masafumi and Okada, Yasuhiro},
  title = {Detailed Calculation of Lepton-Flavor-Violating Muon-Electron Conversion Rate for Various Nuclei},
  journal = {Phys. Rev. D},
  volume = {66},
  pages = {096002},
  year = {2002},
  doi = {10.1103/PhysRevD.66.096002},
  eprint = {hep-ph/0203110},
  archivePrefix = {arXiv},
  note = {Erratum: Phys. Rev. D 76, 059902 (2007)}
}

@article{Doi1985,
  author = {Doi, Masaru and Kotani, Tsuneyuki and Takasugi, Eiichi},
  title = {Double Beta Decay and Majorana Neutrino},
  journal = {Prog. Theor. Phys. Suppl.},
  volume = {83},
  pages = {1--175},
  year = {1985},
  doi = {10.1143/PTPS.83.1}
}

@article{BernardElouadrhiriMeissner2002,
  author = {Bernard, V{\'e}ronique and Elouadrhiri, Latifa and Mei{\ss}ner, Ulf-G.},
  title = {Axial Structure of the Nucleon},
  journal = {J. Phys. G},
  volume = {28},
  pages = {R1--R35},
  year = {2002},
  doi = {10.1088/0954-3899/28/1/201},
  eprint = {hep-ph/0107088},
  archivePrefix = {arXiv}
}

@article{Graf2018,
  author = {Graf, Lukas and Deppisch, Frank F. and Iachello, Francesco and Kotila, Jenni},
  title = {Short-Range Neutrinoless Double Beta Decay Mechanisms},
  journal = {Phys. Rev. D},
  volume = {98},
  number = {9},
  pages = {095023},
  year = {2018},
  doi = {10.1103/PhysRevD.98.095023},
  eprint = {1806.06058},
  archivePrefix = {arXiv},
  primaryClass = {hep-ph}
}

@article{DeppischGraf2020,
  author = {Deppisch, Frank F. and Graf, Lukas and Iachello, Francesco and Kotila, Jenni},
  title = {Analysis of Light Neutrino Exchange and Short-Range Mechanisms in Neutrinoless Double Beta Decay},
  journal = {Phys. Rev. D},
  volume = {102},
  number = {9},
  pages = {095016},
  year = {2020},
  doi = {10.1103/PhysRevD.102.095016},
  eprint = {2009.10119},
  archivePrefix = {arXiv},
  primaryClass = {hep-ph}
}

@article{SimkovicPantis1999,
  author = {{\v S}imkovic, Fedor and Pantis, G. and Vergados, J. D. and Faessler, Amand},
  title = {Additional Nucleon Current Contributions to Neutrinoless Double Beta Decay},
  journal = {Phys. Rev. C},
  volume = {60},
  pages = {055502},
  year = {1999},
  doi = {10.1103/PhysRevC.60.055502},
  eprint = {hep-ph/9905509},
  archivePrefix = {arXiv}
}

@article{Prezeau2003,
  author = {Pr{\'e}zeau, G. and Ramsey-Musolf, M. and Vogel, P.},
  title = {Neutrinoless Double Beta Decay and Effective Field Theory},
  journal = {Phys. Rev. D},
  volume = {68},
  pages = {034016},
  year = {2003},
  doi = {10.1103/PhysRevD.68.034016},
  eprint = {hep-ph/0303205},
  archivePrefix = {arXiv}
}

@article{Cirigliano2021Complete,
  author = {Cirigliano, Vincenzo and Dekens, Wouter and de Vries, Jordy and Hoferichter, Martin and Mereghetti, Emanuele},
  title = {Toward Complete Leading-Order Predictions for Neutrinoless Double-Beta Decay},
  journal = {Phys. Rev. Lett.},
  volume = {126},
  pages = {172002},
  year = {2021},
  doi = {10.1103/PhysRevLett.126.172002},
  eprint = {2012.11602},
  archivePrefix = {arXiv},
  primaryClass = {nucl-th}
}

@article{WirthYaoHergert2021,
  author = {Wirth, R. and Yao, J. M. and Hergert, H.},
  title = {Ab Initio Calculation of the Contact Operator Contribution in the Standard Mechanism for Neutrinoless Double-Beta Decay},
  journal = {Phys. Rev. Lett.},
  volume = {127},
  pages = {242502},
  year = {2021},
  doi = {10.1103/PhysRevLett.127.242502},
  eprint = {2105.05415},
  archivePrefix = {arXiv},
  primaryClass = {nucl-th}
}

@article{DingLiYao2024,
  author = {Ding, Chen-Rong and Li, Gang and Yao, Jiang-Ming},
  title = {Nuclear Matrix Elements of Neutrinoless Double-Beta Decay in Covariant Density Functional Theory with Different Mechanisms},
  journal = {Phys. Lett. B},
  volume = {856},
  pages = {138896},
  year = {2024},
  doi = {10.1016/j.physletb.2024.138896},
  eprint = {2403.17722},
  archivePrefix = {arXiv},
  primaryClass = {nucl-th}
}

@article{YangZhao2025,
  author = {Yang, Y. L. and Zhao, P. W.},
  title = {Next-to-Leading-Order Prediction for the Neutrinoless Double-Beta Decay},
  journal = {Phys. Rev. Lett.},
  volume = {134},
  pages = {242502},
  year = {2025},
  doi = {10.1103/fjy4-wzzq},
  eprint = {2505.24121},
  archivePrefix = {arXiv},
  primaryClass = {nucl-th}
}

@article{ToddEtAl2026,
  author = {Todd, A. and Shickele, T. and Belley, A. and Jokiniemi, L. and Holt, J. D.},
  title = {Ab Initio Short-Range Nuclear Matrix Elements for Neutrinoless Double-Beta Decay},
  journal = {Phys. Rev. Lett.},
  volume = {137},
  pages = {092501},
  year = {2026},
  doi = {10.1103/d2n7-n2c7},
  eprint = {2604.22727},
  archivePrefix = {arXiv},
  primaryClass = {nucl-th}
}

\end{document}